\documentclass{article}
\usepackage[utf8]{inputenc}
\usepackage{authblk}
\usepackage{setspace}
\usepackage[margin=0.5in]{geometry}
\usepackage{graphicx}
\graphicspath{ {./figures/} }
\usepackage{subcaption}
\usepackage{amsmath}
\usepackage{lineno}
\usepackage{hyperref}

\usepackage[backend=biber,style=numeric,sorting=none]{biblatex}
\title{Phase Transitions with Coupled Lasers Array, PhD Research Summary}

\author[1*]{Simon Mahler}

\affil[1]{Department of Physics of Complex Systems, Weizmann Institute of Science, Rehovot 761001, Israel.}
\affil[*]{Email: sim.mahler@gmail.com}

\date{\today}

\begin{document}

\maketitle

\begin{abstract}
Coupled laser arrays exhibit rich and complex physical properties, making them powerful tools for exploring a wide range of phenomena. They enable efficient ground-state optimization of complex landscapes, solve computational problems, reveal topological defects, study coupled oscillators and their universality classes, investigate classical spin systems and complex networks, enhance imaging through scattering media, suppress speckle noise, generate ultra-high-power laser beams, and produce high-resolution shaped beams. Here, I summarize my PhD research on phase-locking large networks of coupled lasers and controlling spatiotemporal lasing modes for rapid speckle reduction.
\end{abstract}

\section{Introduction}
An array of coupled lasers is a network of lasers, each interacting with the others via a well-defined coupling ~\cite{Glova03,Pal20}. The type of coupling can be nearest neighbor coupling where each laser only interacts with its nearest neighbors (i.e. adjacent lasers) to all the way to mean-field coupling where each laser interacts with all the others~\cite{Naresh21,Pal20}. Phase locking of lasers is a consequence of laser coupling and corresponds to a state where all the lasers have the same frequency and the same constant relative phase, leading to a coherent superposition of all the lasers~\cite{Glova03}. Such superposition enhances the overall brightness of the lasers and allows focusing to a sharp spot~\cite{Kong07,Mahler20}. Phase locking of lasers is of interest in diverse investigations, including simulating spin systems~\cite{Nixon13_3,Tamate16,Berloff17,Pal20,Mahler24_spins}, finding ground-state solution of degenerate landscapes~\cite{Nixon13_3,Marandi14,Mahler20}, observing phase transitions and dissipative topological defects~\cite{Pal17,Mahler19,Mahler20_2,Mahler24_spins} and solving hard computational problems~\cite{Marandi14,Tradonsky18}. Phase locking can be achieved in a system of dissipatively coupled laser which brings the system to a stable lasing state of minimal loss~\cite{Mahler20,Nixon13_3,Eckhouse08}. 

Coupled laser arrays are analogous to systems obeying a phase transition between a disordered state to a fully ordered state~\cite{Nixon13_3,Pal20,Berloff17,Tamate16,Marandi14,Mahler20_2}. The disordered state corresponds to uncoupled lasers where each independent laser has a different phase, and the ordered state to coupled lasers where all the lasers have the same frequency and phase. System obeying a phase transition are described by a well-defined modern classification~\cite{Chandler87}. Such classification is rather complex, where some parts remaining theoretically unsolvable and/or experimentally unverifiable. An analogy between coupled lasers arrays and the first and/or the second order phase transitions has been theoretically proposed~\cite{Scott75,Goldstein71,Lifshiftz69,Cohen12,Zamora10}. Since the lasers can be efficiently controlled and manipulated, it would be advantageous to investigate phase transition effects and the modern classification with coupled lasers. 

During my PhD, I experimentally investigated the analogy between arrays of coupled lasers and systems with phase transitions. Specifically, I investigated new coupling techniques \cite{Mahler19_2,Naresh21,Mahler20,Mahler24_spins}, phase transitions with lasers including first and second order transitions \cite{Mahler20_2}, XY model and spin system \cite{Nixon13_3,Mahler20,Pal20,Mahler24_spins}, finding ground-state solution of complex landscape \cite{Mahler20,Pal20,Naresh21}, and topological defects \cite{Mahler19,Pal17,Mahler20_2}. I also investigated the optical properties of the degenerate cavity laser and its spatio-temporal modal structure \cite{Chriki18,Mahler21,Eliezer21,Cao19} leading to applications such as high-resolution beam shaping \cite{Tradonsky21,Roadmap21,Davidson22} and imaging through scattering media \cite{Chriki21,Chriki18,Mahler21}. This report summarizes the investigations and results obtained during my PhD investigations and present future prospects.  

Section~\ref{sec:DCL} introduces the degenerate cavity laser arrangement. Section~\ref{sec:K_methods} presents the investigations and results on different coupling methods for improving phase-locking. Section~\ref{sec:phstrns} presents the investigations and results on phase transitions and topological defects with coupled lasers. Section~\ref{sec:spatmodes} presents the investigations and results on phase-locking spatial modes and the mode structure of the degenerate cavity laser. Section~\ref{sec:conclusion} summarizes the main results obtained during my PhD research. Section~\ref{sec:postdoc} is a short biography with future prospect on laser speckles.  

\section{Degenerate cavity laser}
\label{sec:DCL}
\begin{figure}[h]
    \centering
        \includegraphics[width=0.7\textwidth]{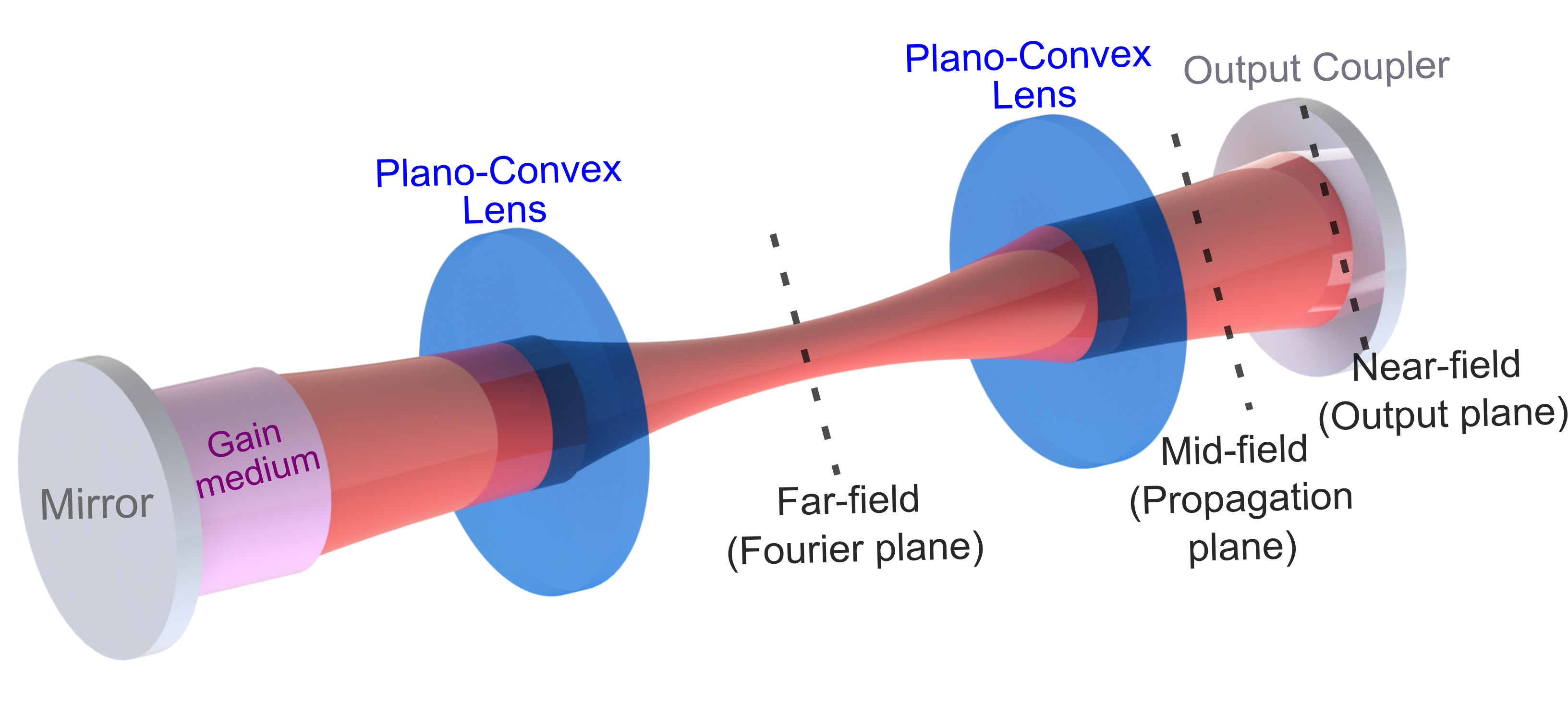}
    \caption{Diagram of the multi-mode degenerate cavity laser (DCL) where spatial and temporal coherences can be controlled by elements in either the near-field, mid-field, or far-field planes, very useful for generating large arrays of lasers and coupling them and for high-resolution, full-field, and fast speckle imaging.}
    \label{fig:1_DCL}
\end{figure}

During most, if not all, the investigations of this report, the degenerate cavity laser (DCL)~\cite{Arnaud69} was used for forming and coupling independent lasers in an array. The DCL arragenment is schematically shown in Figure~\ref{fig:1_DCL}, consisting of a flat back mirror with high reflectivity (generally $>90\%$), a Nd:YAG gain medium attached at the back mirror and lasing at $\lambda=1064nm$, two Fourier lenses in a $4f$ telescope configuration, and a flat front mirror (i.e. output coupler) with $>80\%$ reflectivity.

\subsection{Advantage of the degenerate cavity laser}
The DCL has many advantages compared to typical lasers including self-imaging property, large number of spatial (transverse) modes (up to $320,000$ \cite{Nixon13,Roadmap21}), high number of temporal (longitudinal) modes (up to few hundreds \cite{Chriki18,Mahler21,Eliezer21}), high output power \cite{Mahler19}, control of mode competition and dissipative coupling \cite{Nixon13_3,Eckhouse08,Mahler21,Mahler24_spins}, and relatively easy insertion of intracavity elements (e.g. meta surfaces, mask of holes, gratings, diffusers, lenses, etc). 

These advantages have been successfully exploited in diverse investigations including the formation of arrays of hundreds of independent lasers in an array, which were phase locked~\cite{Nixon13_3,Tradonsky17,Mahler19_2,Mahler20,Mahler20_2,Mahler24_spins}, simulating classical spin systems such as the XY model~\cite{Nixon13_3,Pal20,Mahler24_spins}, investigating the dynamics of complex network of coupled lasers~\cite{Nixon12,Mahler20_2}, solving computational problems by developing a new all-optical approach~\cite{Tradonsky18,Pal20}, investigating dissipative topological defects and Kibble-Zurek mechanism~\cite{Pal17,Mahler19}, efficiently controlling the spatial coherence of the DCL over a large range~\cite{Nixon13,Cao19,Mahler21,Eliezer21}, rapidly reducing speckle contrast by spatio-temporally tuning lasing modes ~\cite{Chriki18,Mahler21,Cao19,Eliezer21}, and imaging and focusing light through intra-cavity scattering elements such as a diffuser~\cite{Nixon13_2,Chriki21,Mahler21}.

\subsection{Near-field, far-field, and mid-field relationships in the DCL}
In a DCL, the near-field and far-field planes are physically related as the focal plane of each other from the spherical lenses. They are also mathematically related through a Fourier transform, meaning that the far-field pattern represents the spatial frequency components of the near-field distribution. The mid-field plane, positioned at a propagation distance $Z$, corresponds to the Fresnel propagator of the near-field, capturing diffraction effects as the beam evolves. Mathematically, all three planes—the near-field, mid-field, and far-field—can be described and numerically simulated using phase synchronization models like the Kuramoto model \cite{Kuramoto84,kuramoto2003chemical} and iterative methods such as the Gerchberg-Saxton algorithm \cite{Gerchberg72}. These simulations are valuable tools for theoretical and numerical predictions. Typically, the Gerchberg-Saxton algorithm is employed to reconstruct the near-field, mid-field, and far-field intensity distributions of a DCL, incorporating parameters such as the gain function and lens focal distance and position. The near-field output represents the laser array’s intensity and phase distribution, while the far-field output corresponds to the coherence function of the laser output. Experimentally,  the near-field, mid-field, and far-field planes can be simultaneously imaged using imaging telescopes \cite{Mahler19_2,Tradonsky21}, enabling direct experimental validation of theoretical models.

\subsection{Detailed experimental arrangement}
In our investigations, the focal length $f$ of the two telescopic lenses in Fig.~\ref{fig:1_DCL} ranged from $f=[10$ to $100]$ cm. The $4f$ telescope images the mirror plane onto the output coupler plane, ensuring that any hole in the mask is precisely imaged onto itself after a single round trip (self imaging). Thus, each hole in the mask serves to form an independent laser~\cite{Mahler19_2,Tradonsky17}. The fields adjacent to the mirrors of the DCL are denoted near-field. The field midway between the lenses is denoted far-field and is analytically equivalent to the Fourier transform of the near-field \cite{vonBieren71}. The near-field and far-field planes are physically accessible and controllable in our DCL. Specifically, one can place elements at these planes (e.g. a near-field mask of holes or spatial light modulator) adding constraints and breaking the mode degeneracy in the DCL, so as to provide control on many degrees of freedom. 

To reduce the effects of thermal lensing, the gain medium is pumped with quasi-CW $100$ $\mu$s long pulses Xenon flash lamps operating at $1$ Hz. Note that more recently, the gain medium can be pumped with a high-power laser diode. The diameter of the gain medium is typically $0.95$ cm. Its cross section is large compared to the diffraction spot size of the telescope, ensuring a large number $N$ of spatial (transverse) lasing modes inside the DCL (without a mask), up to $N=320,000$ \cite{Nixon13}. The number of lasing modes $N$ can be controlled from $1$ to $320000$ with less than a $50\%$ reduction in the total output power \cite{Nixon13}. Therefore, the DCL can serve as a spatially coherent or a highly spatially incoherent source \cite{Chriki18,Cao19,Nixon13}.

\section{Coupling methods for improved phase-locking}
\label{sec:K_methods}
Phase locking of laser arrays can be achieved with dissipative coupling that drives the system to a stable state of minimal loss, which is the phase locked state \cite{Pal17,Eckhouse08,Nixon13_3}. Dissipative coupling involves mode competition whereby modes of different losses compete for the same gain, and only modes with the lowest loss survive and are amplified by the gain medium \cite{Eckhouse08,Nixon13_3,Glova03}. Accordingly, by inserting amplitude and phase linear optical elements into a laser cavity that minimize the loss of the phase locked states, it is possible to achieve phase locking with mode competition \cite{Glova03,Kong07,Mahler19_2,Zhou04}. While phase locking with such linear optical elements has yielded many exciting results \cite{Eckhouse08,Nixon13_3,Pal17,Tradonsky18,Tradonsky17,Zhou04,Guillot11,Sivaramakrishnan15,Nixon13_2}, it suffers from several inherent limitations. For example, it is very sensitive to imperfections, such as positioning errors, mechanical vibrations, size and orientation of the lasers, thermal effects and various types of aberrations associated with these intra-cavity elements. Moreover, in many cases, especially for spin simulations and computational problem solving \cite{Marandi14,Tradonsky18,Nixon13_3}, there are two or more states with nearly degenerate minimal loss that cannot be distinguished from each other. Accordingly, it is difficult to phase lock disordered arrays of lasers in the in-phase state with these coupling techniques. Other complicated array geometries such as Kagome and triangular will inherently suffer from inconsistencies between nearest and next-nearest neighbors coupling phase \cite{Pal20,Nixon13_3,Chalker92}. Even a simple square array geometry requires precise setting and positioning of the coupling elements that could be optimized only for a uniform and specific separation between the lasers \cite{Mahler20,Tradonsky17}.

This section presents various coupling methods explored during my PhD to enhance and develop new configurations of phase-locked laser arrays. First, far-field coupling with lenses and diffractive optical elements provides versatile configurations. Second, nonlinear coupling using a far-field saturable absorber improves sensitivity to loss differences and helps identify unique solutions in near-degenerate cases. Third, Gaussian coupling with a far-field Gaussian aperture enables stable in-phase locking regardless of array geometry while suppressing internal phase structures for high-quality laser output. Finally, mid-field coupling with a mask enables the realization of the classical XY Hamiltonian, allowing the study of topological defects and potentially Hofstadter physics.

\subsection{Controlled far-field couplings \cite{Mahler19_2}}
\label{sec:FFK}
Many different techniques for coupling lasers have been developed, involving evanescence~\cite{Shirakawa02,Fridman07}, anti-guiding~\cite{Botez91}, mirrors~\cite{Fridman10}, Fourier diffraction, fractional Talbot diffraction~\cite{Amato89,Tradonsky17} and Gaussian coupling~\cite{Naresh21}. Unfortunately, the coupling with these techniques cannot be readily controlled, often suffer from spurious and undesired coupling, or depend on the specific configuration geometry of the lasers in an array. This is particularly true for an array of lasers formed in a DCL where the coupling between individual lasers is done with Talbot diffraction and/or Fourier diffraction aperture. In the past, fractional Talbot diffraction was exploited to efficiently couple up to $1500$ lasers formed in a DCL~\cite{Nixon13_3,Tradonsky17}, but it was found that the coupling between the lasers depends on accurate control of the configuration, geometry, and period of the array.

\begin{figure}[h!]
\centering\includegraphics[width=0.65\textwidth]{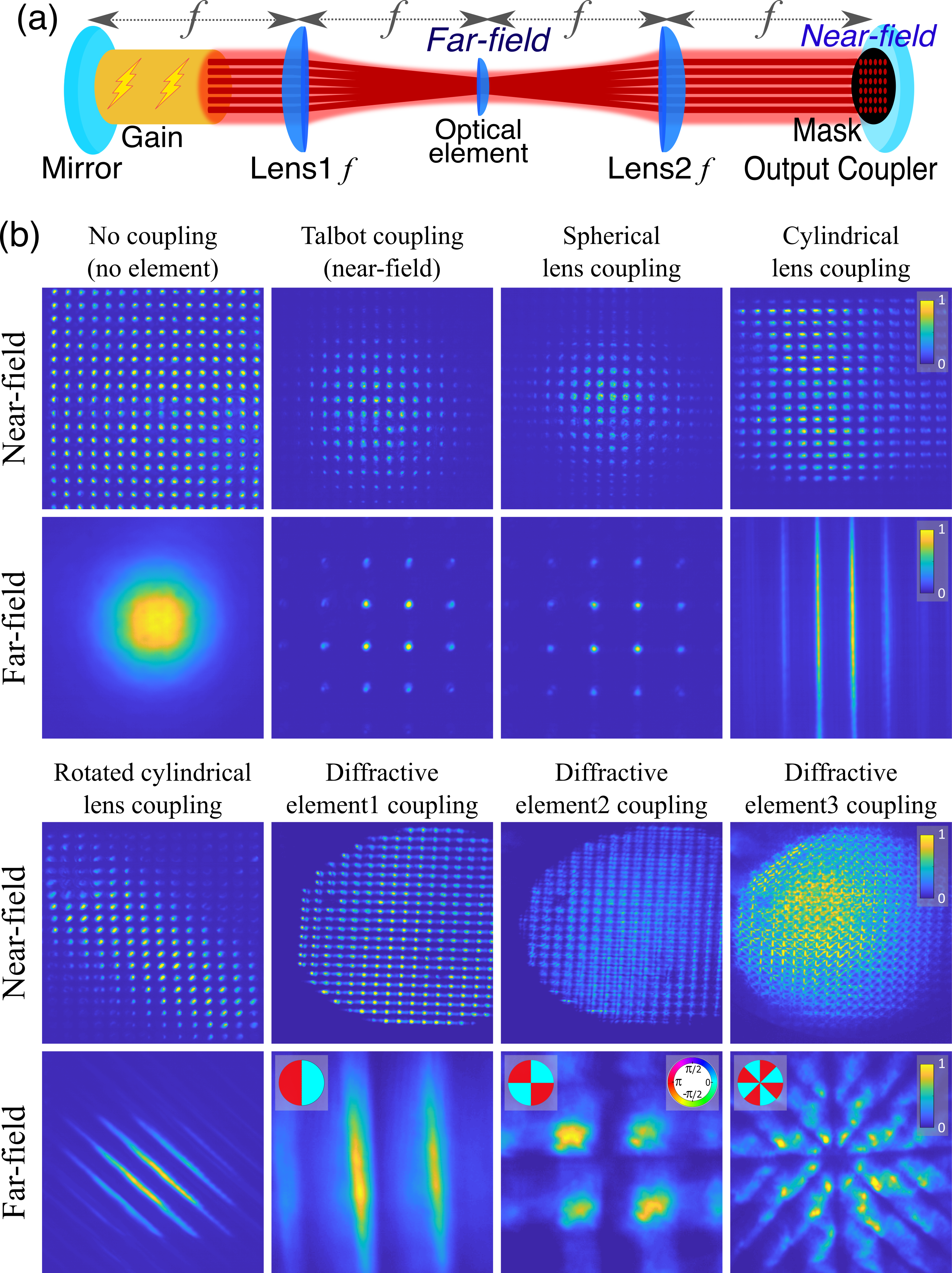}
\caption{Experimental arrangement and results for versatile far-field coupling. (a) DCL arrangement with an optical element located on the far-field for coupling lasers. (b) Coupling results with different optical elements by showing the near-field and far-field intensity distributions of the lasers.}
\label{fig:5_ff_K}
\end{figure}

In this section, we present a relatively new technique to couple lasers, by means of intra-cavity optical elements, where the coupling is versatile and relatively easy to control. The experimental arrangement and typical results are presented in Fig.~\ref{fig:5_ff_K} \cite{Mahler19_2}. 

The intra-cavity optical elements in the far-field plane of the DCL are inserted in order to control coupling between lasers in an array regardless of the geometry and configuration of the array. When the intra-cavity element is a spherical lens, we experimentally and numerically found that the coupling results are similar to those obtained with the Talbot diffraction technique, but are easier to obtain and there is greater versatility. The new coupling technique not only mimics the Talbot diffraction, but also offers control on the selection of the lasers to couple. Specifically, one-dimensional coupling in a two-dimensional square array of lasers is achieved by replacing the spherical lens with a cylindrical lens. By rotating the cylindrical lens, it is possible to choose the orientation of the lasers to couple. By replacing the spherical lens with a binary phase element (BPE), it is possible to select lasers to couple and to control the coupling in a variety of array geometries~\cite{Cheung08,Mahler19_2}. Specifically, by properly designing grating arrays and BPEs, it is possible to obtain versatile and controlled coupling of lasers when placed in the far-field plane of a DCL, as the BPE could impose a far-field intensity distribution, having the ability to couple the lasers in an array. Following on this new far-field coupling technique, we next investigated nonlinear coupling with a far-field saturable absorber and all-positive-coupling with a far-field Gaussian aperture.

\subsection{Nonlinear coupling \cite{Mahler20}}
\begin{figure}[!ht]
\centering
\includegraphics[width=0.65\textwidth]{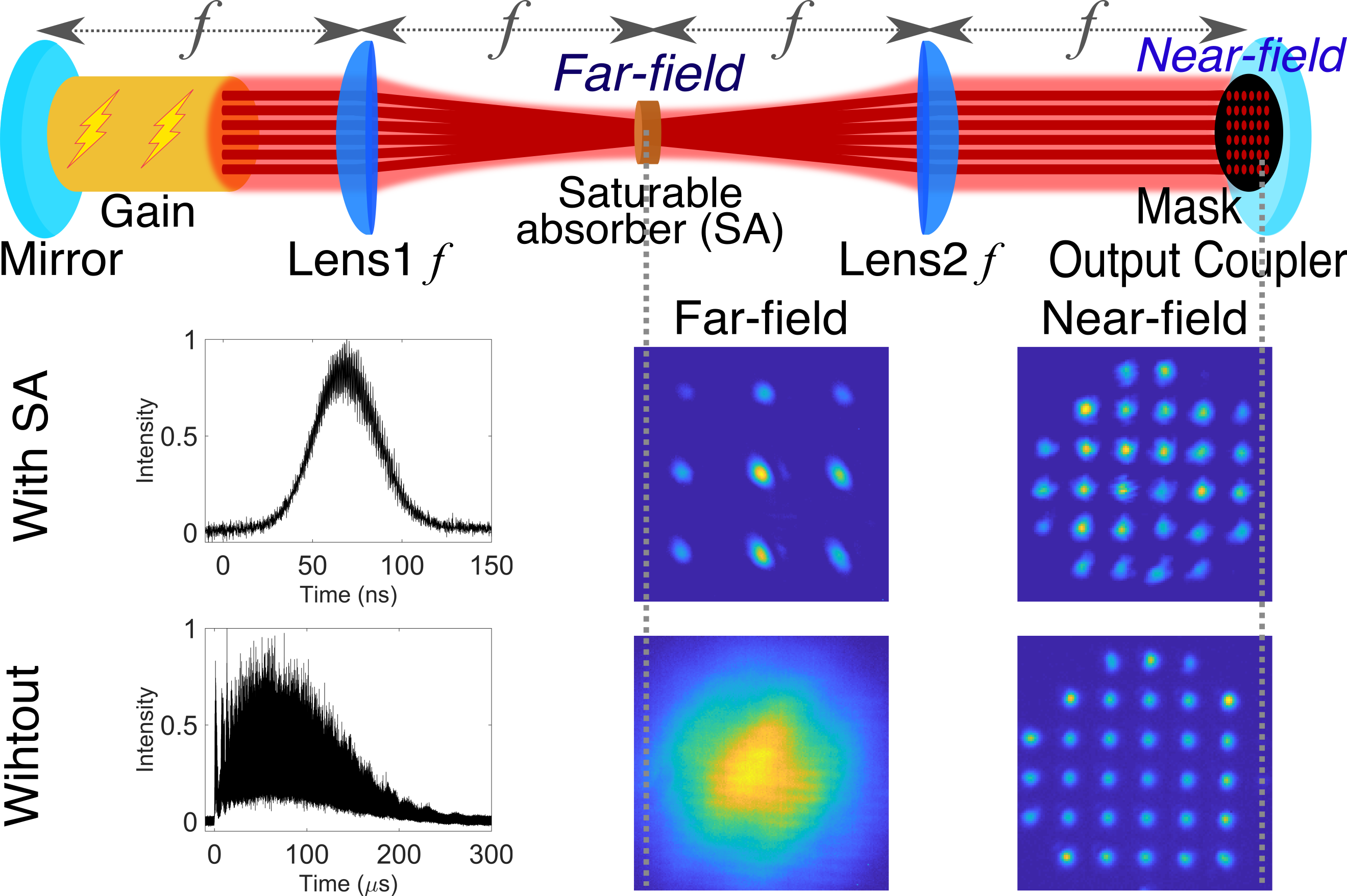}
\caption{Degenerate cavity laser arrangement with a far-field saturable absorber for nonlinearly coupling lasers in a square array. The time evolution of the laser array intensity is reduced from $200$ $\mu$s to $100$ ns by the insertion of the saturable absorber, indicating temporal Q-switching. The broad Gaussian far-field intensity distribution without the saturable absorber indicates that there is no phase relation between the lasers whereas the sharp peaks with the saturable absorber indicate near-perfect phase locking of the entire array.}
\label{fig:2_SA_K}
\end{figure}

To achieve nonlinear (spatio-temporal) coupling, we resorted to a far-field DCL intra-cavity saturable absorber \cite{Mahler20}. A saturable absorber is a nonlinear optical element that blocks light until it saturates, where its optical loss decreases sharply \cite{Hercher67}. Our experimental arrangement, comprising of a DCL with a far-field Cr:YAG saturable absorber, is presented in Fig.~\ref{fig:2_SA_K}. As shown in the figure, without a saturable absorber, the lasing dynamics is complex with strong oscillations and a long pulse duration of about $200$ $\mu$s. With a saturable absorber, the lasing dynamics is a Gaussian-enveloppe pulse of duration of 100 ns indicating temporal Q-switching \cite{Siegman86}. The total output powers in both cases were similar, thereby the peak pulse power with a saturable abosrber being significantly higher than without. 

The far-field and near-field intensity distributions of the DCL in Fig.~\ref{fig:2_SA_K} results indicate that the about $30$ lasers were temporally Q-switched and spatially phase locked with a saturable abosrber. Indeed, the near-field intensity distributions of the square array of lasers without and with the saturable absorber are essentially the same, while the far-field intensity distributions differ dramatically, indicating different spatial coherences. Specifically, the broad Gaussian in the far-field intensity distribution without the saturable absorber indicates no phase relation between the different lasers in the array \cite{Kong07,Tradonsky17,Mahler19_2}. On the other hand, the sharp peaks in the far-field intensity distribution with the saturable absorber indicate in-phase locking of the about $30$ lasers in the array \cite{Kong07,Tradonsky17,Mahler19_2}. These high sharp intensity peaks increase the saturation  of the saturable absorber and thereby minimize loss. This minimal nonlinear loss configuration is thereby selected by the mode competition, explaining phase locking mechanism of the lasers by the saturable absorber. Here we show a case of 30 lasers, but more lasers can be phase locked with the saturable absorber.

We also investigated whether adding nonlinear coupling to linearly coupled lasers can improve the convergence to the lowest loss phase locked state (e.g. out-of-phase state), when an additional state (e.g. in-phase state) with nearly identical but slightly higher loss is present, see Ref. \cite{Mahler20}. The arrangement with nonlinear coupling (i.e. with the saturable absorber) was found to be $25$ times more sensitive to loss differences and converged $5$ times faster to the lowest loss phase locked state than with linear coupling \cite{Mahler20}, providing a unique solution to problems that have several near-degenerate solutions.

\subsection{Gaussian coupling \cite{Naresh21}}
\label{sec:GK}
\begin{figure}[h!]
\centering\includegraphics[width=0.6\textwidth]{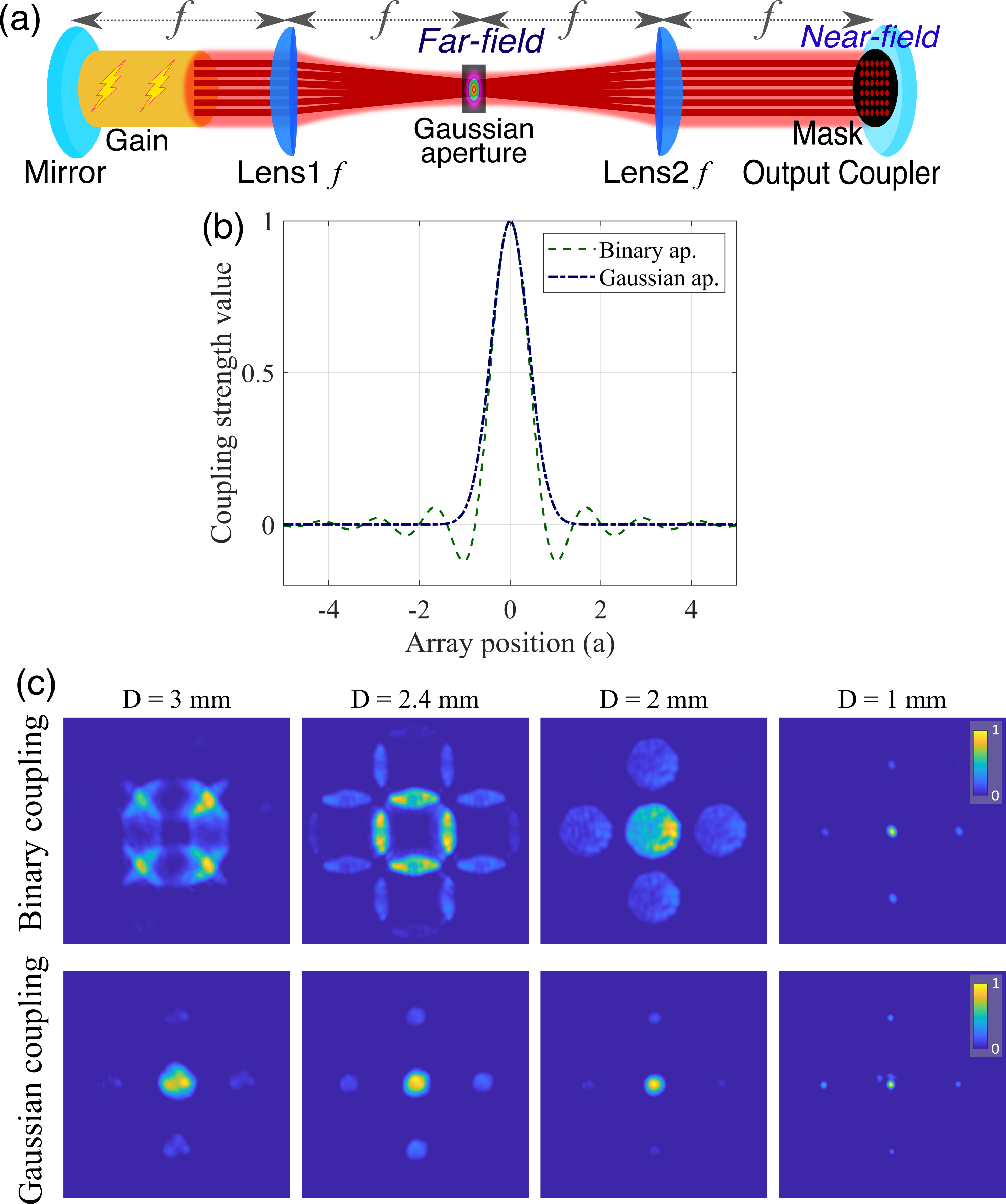}
\caption{Experimental arrangement and results for forming a Gaussian coupled array of lasers. (a) A Gaussian aperture is inserted at the far-field (Fourier) plane of a degenerate cavity laser and a mask of holes is inserted at the near-field plane. (b) Binary and Gaussian apertures coupling functions. With a binary aperture, the sign, range, and strength of the coupling depends on the period, orientation and size of the lasers. With a Gaussian aperture, the coupling is always positive. (c) Binary versus Gaussian couplings. Detected far-field intensity distributions of the lasers in a square array for different diameters $D$ of intra-cavity binary apertures and Gaussian apertures. As evident, with binary apertures, the lasers phase lock with either positive or negative or other coupling, depending on the aperture size; with Gaussian apertures, the lasers always lock in-phase, regardless of the size of the aperture.}
\label{fig:4_G_K}
\end{figure}

In this section, a different and relatively simple coupling technique based on a Gaussian coupling function is presented, Figs.~\ref{fig:4_G_K}(a) and (b), where the Gaussian coupling function is always real and positive, ensuring that all the lasers are always positively coupled regardless of the array geometry, position, orientation, period or size, thereby removing frustration associated with other coupling techniques where the sign of the coupling changes with the distance and with the array geometry \cite{Naresh21}. In Fig.~\ref{fig:4_G_K}(c), about $90$ independent lasers formed in a degenerate cavity laser were efficiently in-phase locked by a Gaussian aperture that was inserted in the far-field plane. Additionnaly, steady in-phase locking of lasers  was obtained also for Kagome and random arrays of lasers, even in the presence of near-degenerate solutions, geometric frustration or superimposed longitudinal modes \cite{Naresh21}. Whereas such in-phase locking with other coupling methods, e.g. with the Talbot coupling method, significant losses are observed and the method requires a precisely tuned and extremely small aperture to avoid superposition of degenerate lowest loss states.  Finally, internal phase structures of the lasers were suppressed, so as to result in pure Gaussian laser outputs with uniform phase and improved overall beam quality \cite{Naresh21}. 

Therefore, Gaussian coupling is a simple and robust approach for steady in-phase locking of lasers. The coupling function is always positive as opposed to other coupling functions, where frustration and phase oscillations can arise due to the changes of sign in the coupling function. The range of the coupling as well as the number of lasers that are phase locked can be gradually controlled from no coupling to mean-field (all to all) coupling by changing the diameter of the Gaussian aperture \cite{Naresh21}. Steady in-phase locking of lasers can also be achieved in a Kagome and quasi-random laser arrays. Such in-phase locking can be achieved within less than $100$ ns and with a high power density sharp spot in the far-field \cite{Mahler20,Naresh21}. In addition to providing coupling between different lasers in the array, the far-field Gaussian aperture also serves as a spatial filter, ensuring that each laser in the array is a pure Gaussian mode with no internal structure \cite{Naresh21}. Such pure single-mode Gaussian lasers are advantageous for many phase locking applications and in particular in spin simulators and solvers \cite{Pal20,Mahler24_spins}.  

\newpage
\subsection{Mid-field coupling \cite{Mahler24_spins}}
\label{sec:MFK}
In all of the above methods, phase locking of lasers was achieved via dissipative coupling where the coupling is purely real~\cite{Mahler20,Nixon13_3,Mukherjee17} or via dispersive coupling where the coupling is imaginary. Recently, complex coupling was considered by combining dissipative and dispersive couplings~\cite{Leefmans21,Miri19}. Such combined coupling, where both the strength and the phase can be controlled, leads to the investigation of many unexplored couplings, systems and configurations such as anti-symmetrical coupling of lasers, where the phase of the coupling is varied while keeping the coupling strength constant, arbitrary coupling, the ability to break time-reversal symmetry and the formation of an artificial gauge field.  In this section, we present anti-symmetrical coupling between lasers, implemented by the means of a ring degenerate cavity laser and a mid-field coupler mask \cite{Mahler24_spins}. 

\begin{figure}[!ht]
\centering\includegraphics[width=0.85\textwidth]{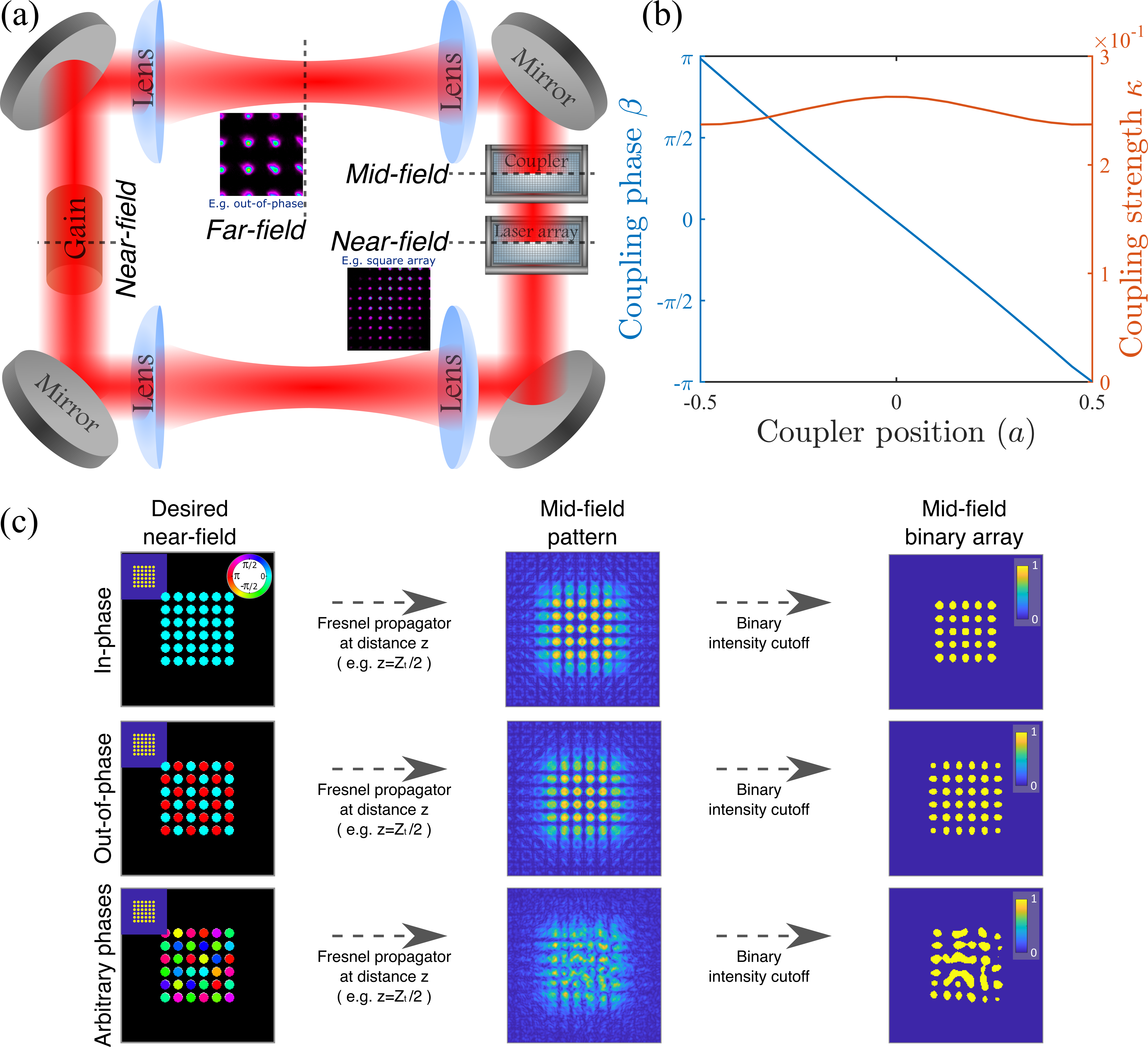}
\caption{Arrangement and procedure for implementing anti-symmetrical phase-coupling of lasers. (a) Conceptual experimental arrangement of a ring digital degenerate cavity laser for forming and coupling lasers in an array. (b) Phase (blue curve) and strength (orange curve) of the coupling between a pair of nearest neighbors lasers as a function of the coupler position where $a$ is the period of the array. (c) Process for designing a coupler array to be inserted in the mid-field plane.}
\label{fig:6_MF_K_method}
\end{figure}

The experimental arrangement is based on a ring digital degenerate laser cavity (DDCL) \cite{Tradonsky21,Mahler24_spins,Davidson22}. The ring DDCL, schematically shown in Fig.~\ref{fig:6_MF_K_method}(a), is a laser cavity that can support a large number of spatial lasing modes and is robust to misalignments and imperfections \cite{Mahler21}. In our ring DDCL, the laser light propagates unidirectionally inside the cavity. A mask array of holes, inserted at the near-field plane is used to form an array of lasers with period $a$ and controls the intensity of the DDCL. A coupler, inserted at a distance $z=Z_{t}/2$ from the near-field (namely a mid-field plane) with $Z_{t}=a^{2}/\lambda$ the Talbot distance of the array and $\lambda=1064$ nm the laser light wavelength, couples the lasers in the array. The near-field, mid-field, and far-field planes of the DDCL are all simultaneously imaged and detected onto a camera. 

We performed numerical calculations on the behavior of the coupling between the lasers for such a ring DDCL arrangement. The results, for a pair of nearest neighbor lasers, are presented in Fig.~\ref{fig:6_MF_K_method}(b), where we show the coupling phase and strength as a function of the coupler horizontal position for a line array of $15$ lasers. The position zero of the coupler corresponds to the position where in-phase locking of lasers (positive coupling) is achieved. The position $\pm a/2$ of the coupler corresponds to the position of the coupler where out-of-phase locking of lasers (negative coupling) is achieved. As evident, the coupling strength is symmetrical around the zero position of the coupler while the coupling phase is anti-symmetrical. The coupling strength is maximal at the zero position of the coupler and slightly decreases as the coupler position deviates from zero position. The coupling phase is zero at the zero position of the coupler and positively (negatively) increases to $\pi$ as the coupler position increases (decreases). Such anti-symmetrical coupling can be used for implementing arbitrary phase locking of lasers and also for implementing artificial gauge field in the laser array, see Section~\ref{sec:Gfield}.

The process for determining the mid-field coupler array that can couple an array of lasers is shown in Fig.~\ref{fig:6_MF_K_method}(c). First, a near-field with the desired intensity and phase distributions is initialized. Then, it is propagated by using a Fresnel propagator to a distance $z$ (diffracted mid-field), smoothed by applying a Gaussian filter and converted into a binary array of transmissions of either $0$ or $1$ by applying an intensity cutoff. In our DDCL, the near-field intensity distribution was controlled via a spatial light modulator (SLM) and the near-field phase distribution was controlled via a coupler array that was a physical binary mask of holes inserted in the mid-field plane. The coupler array can also be implemented via a transmissive SLM inserted in the mid-field plane. With such process, any desired phase distribution of the lasers can be implemented.

\begin{figure}[!ht]
\centering\includegraphics[width=0.55\textwidth]{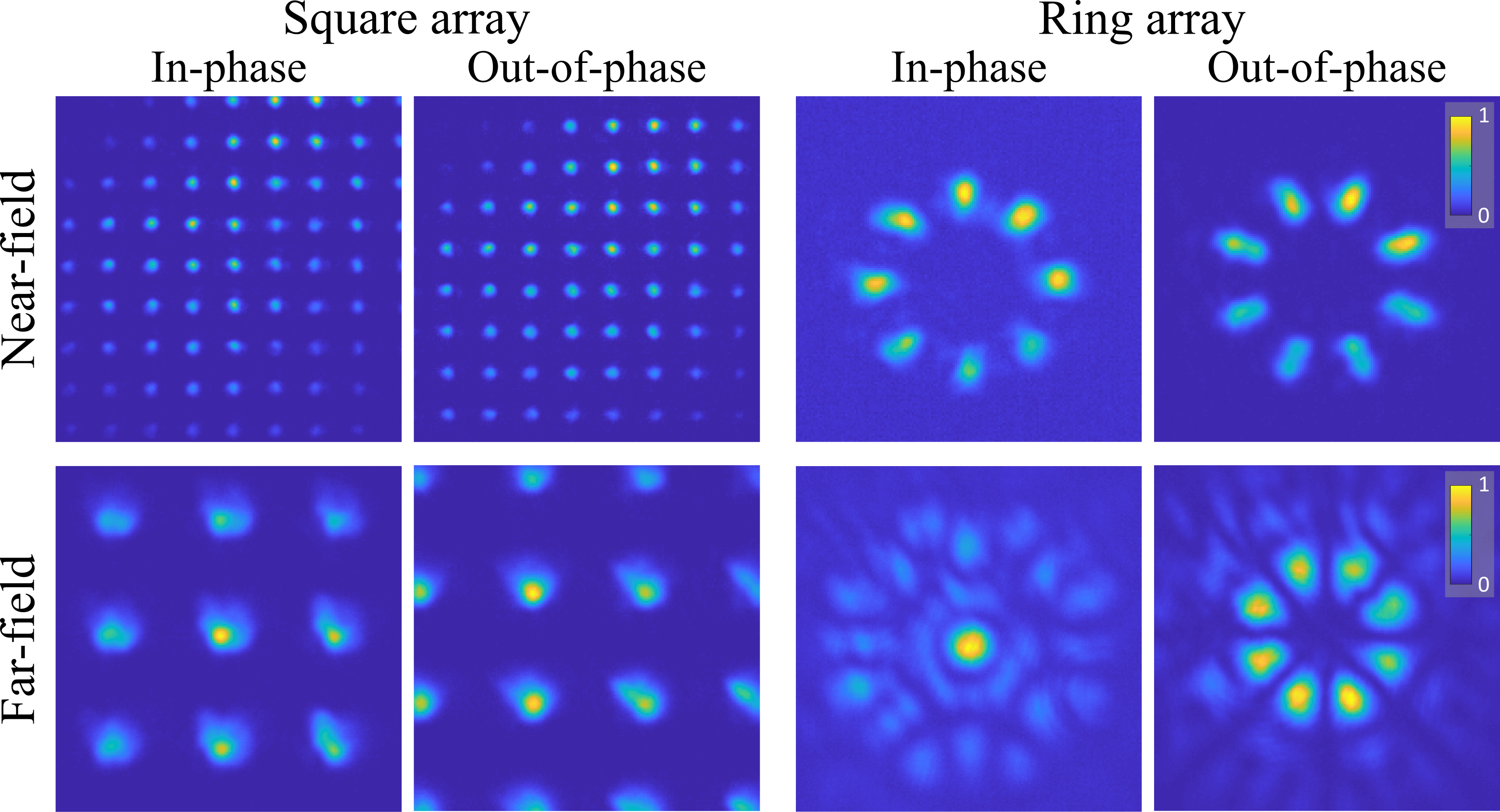}
\caption{Phase locked lasers in the in-phase and in the out-of-phase locking states in a square array and a ring array of lasers that are coupled with a mid-field coupled array.}
\label{fig:7_sq_rng_MFK}
\end{figure}

Figure~\ref{fig:7_sq_rng_MFK} shows represenative examples of phase locked lasers in the in-phase and in the out-of-phase locking states, for a square array and a ring array of lasers. As evident, the bright spots in the far-field indicate that phase-locking between the lasers was efficiently implemented. With such anti-symmetrical coupling, one can implement arbitrary phase locking of lasers and also artificial gauge field to study topological photonics, see Section~\ref{sec:Gfield} and Ref.!\cite{Mahler24_spins}.

\section{Phase transitions and topological defects with coupled lasers}
\label{sec:phstrns}
In this chapter, investigations on phase transitions and topological defects with coupled lasers are presented. Section~\ref{sec:crowdsynch} presents crowd synchronization effect with coupled lasers where a first-order transition to synchronization is observed~\cite{Mahler20_2}. Section~\ref{sec:percolation} presents percolation with coupled lasers where a second-order phase transition is observed. Section~\ref{sec:fairsamp} presents fair sampling of XY spin Hamiltonian with coupled lasers~\cite{Pal20}. Section~\ref{sec:KZM} presents results about Kibble-Zurek mechanism and dynamics of topological defects in the Kuramoto model~\cite{Mahler19}. Finally, Section~\ref{sec:Gfield} shows how generating a synthetic gauge field with mid-field coupling method can be useful for investigating topological defects and Hofstadter physics.

\subsection{Crowd synchrony and first-order transitions \cite{Mahler20_2}}
\label{sec:crowdsynch}
\begin{figure}[!ht]
\centering
\includegraphics[width=0.6\textwidth]{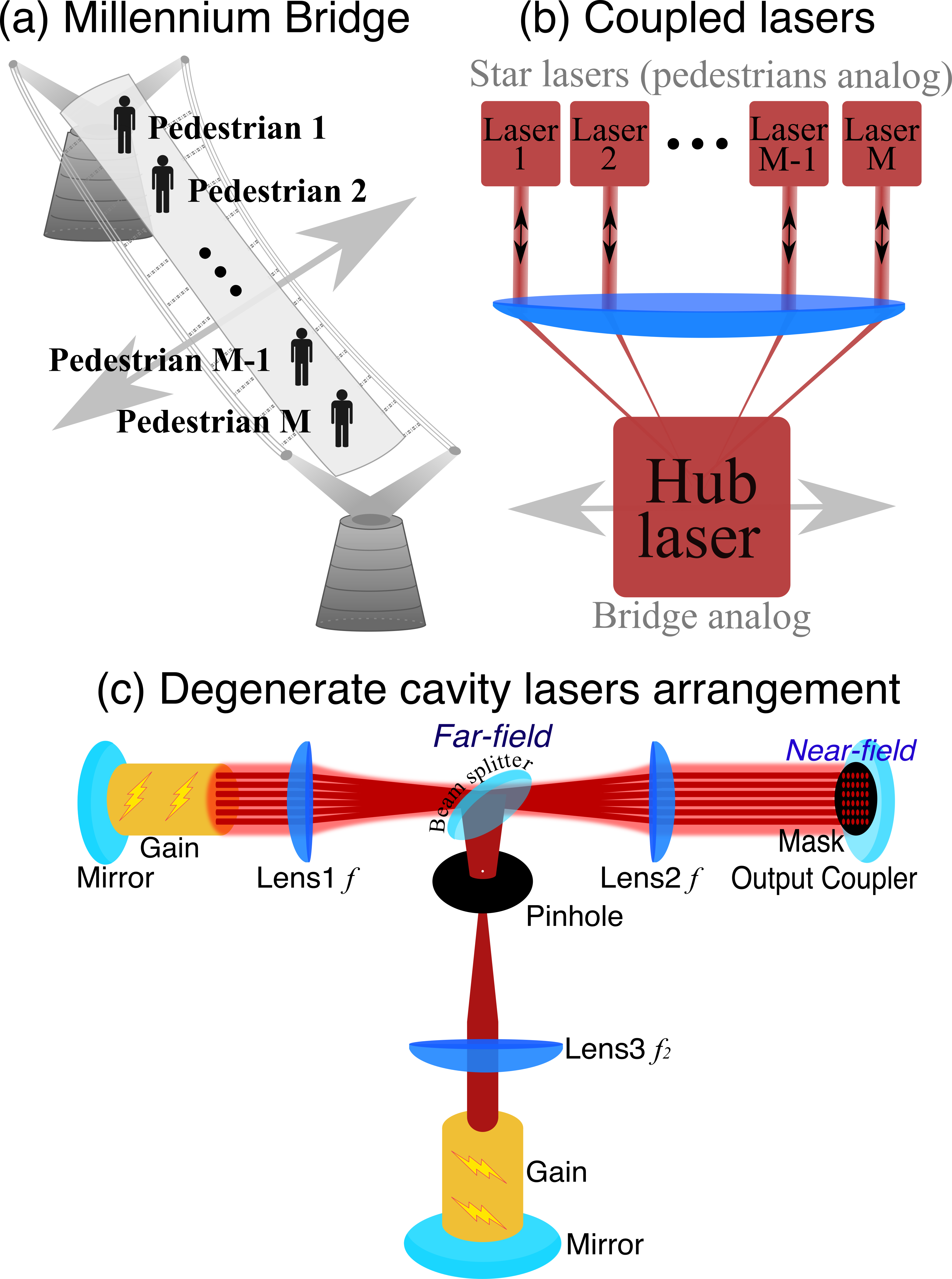}
\caption{Configurations for observing crowd synchrony. (a) Millennium Bridge, (b) and (c) coupled lasers. (c) Experimental arrangement with degenerate cavity lasers (DCLs). The star DCL forms independent (uncoupled) lasers and the hub DCL cavity couples them with long range. A variable size aperture placed adjacent to the mask of holes controls the number of lasers.}
\label{fig:8_exp_skecth}
\end{figure}

Crowd synchrony was first observed on the pedestrian Millennium Bridge in London, that had to be partially closed on its opening day in $2000$ due to uncontrolled strong lateral oscillations. Shortly after, it was noted that the strength of the lateral oscillations can be decreased by reducing the number of pedestrians crossing the bridge \cite{Dallard01}. Further investigations revealed that the pedestrians, each walking at different frequency (and phase), caused small lateral oscillations to the bridge, which in turn caused the pedestrians to sway in step in order to retain balance, dramatically amplifying the lateral oscillations of the bridge and synchronizing the pedestrians. In 2005, the effect was fully modeled as crowd synchronization, where the lateral oscillations of the bridge were explained as a combination of oscillations and synchronization (Fig.~\ref{fig:8_exp_skecth}(a)) \cite{Strogatz05,Eckhardt07}. Crowd synchrony was modeled by the synchronization of different and independent oscillators interacting each other via a common intermediary, where the synchronization critically depends on the number of oscillators (e.g. pedestrians). In 2010, a theoretical model to demonstrate crowd synchrony in delay coupled lasers was proposed and developed by using the laser rate equations (Fig.~\ref{fig:8_exp_skecth}(b)) \cite{Zamora10}. 

In this section, we confirm for the first time the theoretical model by experimentally demonstrating crowd synchrony with coupled lasers \cite{Mahler20_2}. It is achieved by coupling two degenerate cavity lasers (DCLs),  Fig.~\ref{fig:8_exp_skecth}. The first (star) DCL forms independent lasers with (almost) uniform intensities but different phases and frequencies. The second (hub) DCL, operating below lasing threshold, efficiently couples and phase locks them. As in crowd synchrony, below a critical number of lasers, the lasers remain uncoupled while above most, if not all, the lasers are phase locked. As theoretically predicted, the synchronization of the lasers follows a first-order-like transition as a function of their number. The critical numbers of lasers depends on the coupling between the two cavities and follows a power law $M_{c}\propto K^{\nu}$ with negative exponent $\nu=-2.4$ in good agreement with the theoretically predicted exponent $\nu_{th}=-2.2$ \cite{Zamora10}.  

The spatial coherence of the two DCLs was characterized by measuring the common near-field and far-field intensity distributions and then extracting the coherence peak ratio and percentage of lasers phase-locked \cite{Mahler20_2}. The coherence peak ratio measures the intensity ratio between the phase-locking peaks in the far-field and the background level, similarly to the total coherent intensity used in \cite{Zamora10}. A high coherence peak ratio indicates that most of the lasers are synchronized in-phase and a low coherence peak ratio indicates that most of the lasers are not synchronized. The percentage of lasers phase-locked quantifies the phase locking range and is calculated by dividing the separation distance between phase-locking peaks to its width \cite{Naresh21}. 

\begin{figure}[!ht]
\centering
\includegraphics[width=0.6\textwidth]{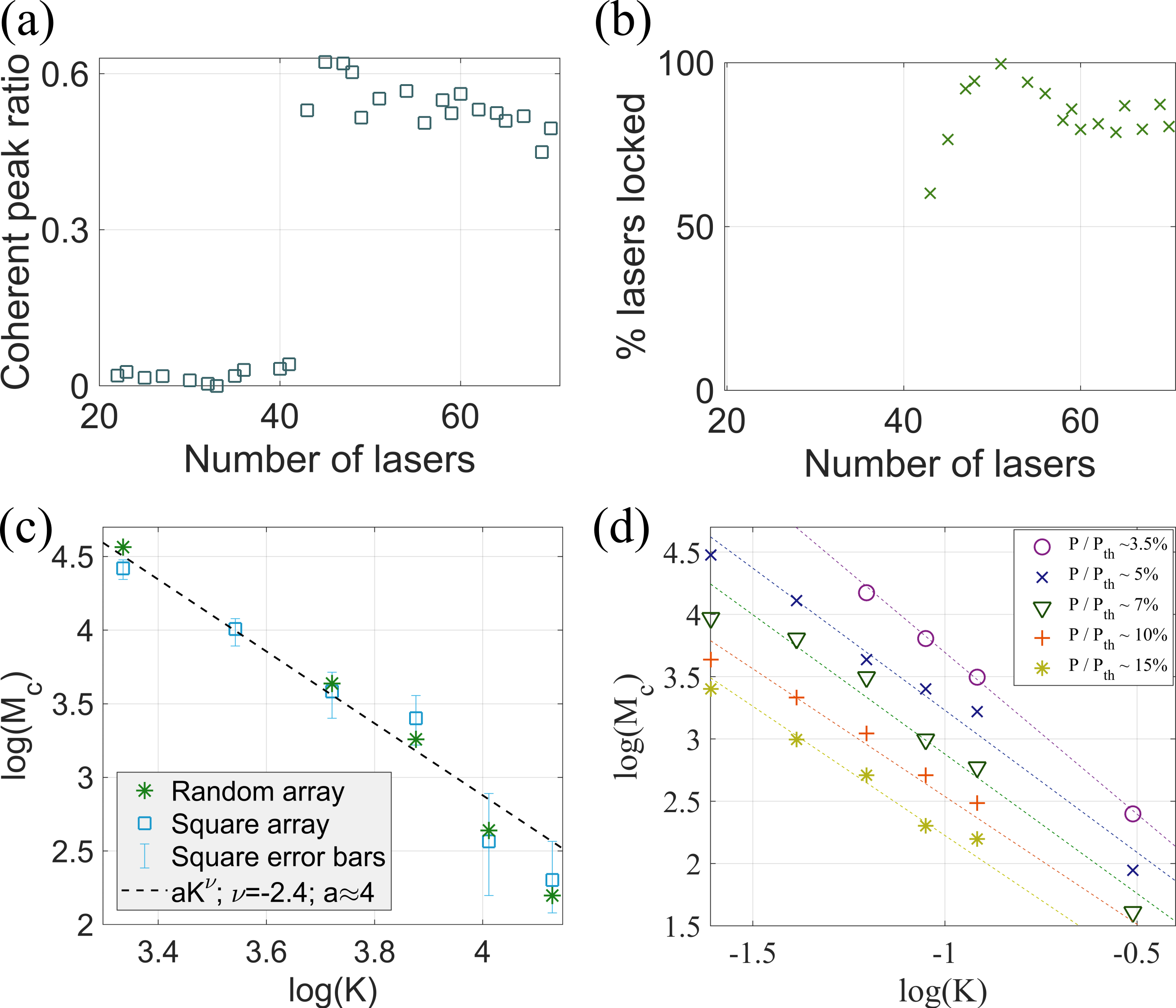}
\caption{Synchronization of the lasers as a function of the number of lasers in a square array. (a) Coherence peak ratio as a function of the number of lasers and (b) percentage of lasers phase locked as a function of the number of lasers. The synchronization of the lasers follows a first-order-like transition as the number of lasers increases where the critical number of lasers is $M_{c}=43$. (c) Measured critical number of lasers $M_{c}$ as a function of the coupling $K$ between the two DCLs for two different array geometries (square and random). (d) Numerical simulation where the pump ratio (pump over pump threshold) of the hub laser was tuned. The critical number of lasers decreases as the coupling increases and follows a power law $M_{c}\propto K^{\nu}$ with $\nu_{exp}=-2.4$ (numerically $\nu_{num}=-2.2$). }
\label{fig:9_M_c}
\end{figure}

The results in Fig.~\ref{fig:9_M_c}(a) and (b), obtained for a square of lasers, indicate that the synchronization of the lasers critically depends on their number. Below a critical number ($M_{c}=43$ here), the lasers remain uncoupled, as indicated by a low coherence peak ratio and a $0\%$ percentage of lasers phase-locked. Thereby, the star DCL is lasing but not the hub DCL, i.e. the lasers from the star DCL do not have enough energy to pump and force the hub DCL to lase, as in the Millennium Bridge effect where there are no strong lateral vibration below a critical number of pedestrians. Above $M_{c}$, a sharp increase occurs in the coherence peak ratio to $0.6$, indicating that both (star and hub) DCLs are lasing and that most of the lasers are phase locked. The lasers from the star DCL have enough energy to pump and force the hub DCL to lase, where the lasers are coupled (via the Fourier pinhole). 

We also repeated the measurements in Figs.~\ref{fig:9_M_c}(a) and (b) with a random array of lasers and similar results that the one with the square array were observed, see~\cite{Mahler20_2}.

Next, we determined the critical number of lasers $M_{c}$ as a function of the coupling $K$ between the two DCLs for the two different array geometries (square and random). The coupling $K$ corresponds to the percentage of laser light propagating from the star DCL to the hub DCL (via the polarizing beam splitter). This percentage can be tuned by rotating the polarizer (quarter wave-plate) in the experimental arrangement of Fig.~\ref{fig:8_exp_skecth}(c) from its fast axis, where from Malus's law, the coupling is calculated as $K=sin^{2}(4\theta)$, with $\theta$ is the angle between the horizontal axis and the fast-axis of the quarter wave-plate. The results are presented in Fig.~\ref{fig:9_M_c}(c) and (d), showing the $log$ of the critical number of lasers as a function of the $log$ of the coupling. As evident, the critical number of lasers decreases as the coupling increases. It seems that the array geometry does not influence the behavior of the critical number of lasers. As indicated by the fit $aK^{\nu}$, the critical numbers of lasers follows a power law $M_{c}\propto K^{\nu}$ as a function of the coupling with a negative exponent $\nu_{exp}=-2.4$ (numerically $\nu_{num}=-2.2$), that is in good agreement with the theoretically predicted exponent $\nu_{th}=-2.2$ in~\cite{Zamora10}. We also performed additional numerical investigations on the scaling exponent where the pump of the hub DCL was varied, as shown in Fig~\ref{fig:9_M_c}(d). 

To conclude, we developed an arrangement, in which we incorporated two coupled degenerate cavity lasers, one with an array of lasers to form independent lasers and the other with a Fourier spatial filter pinhole to couple them. We experimentally demonstrated crowd synchrony with coupled lasers and showed that the synchronization of the lasers depends on the number of lasers and follows a first-order-like transition. The critical number of lasers as a function of the coupling between the two cavities followed a power law with exponent $\nu=-2.4$ similar to the theoretically predicted one $\nu_{th}=-2.2$.

\subsection{Nonlinear percolation and second-order transitions}
\label{sec:percolation}
\begin{figure}[!ht]
\centering
\includegraphics[width=0.6\textwidth]{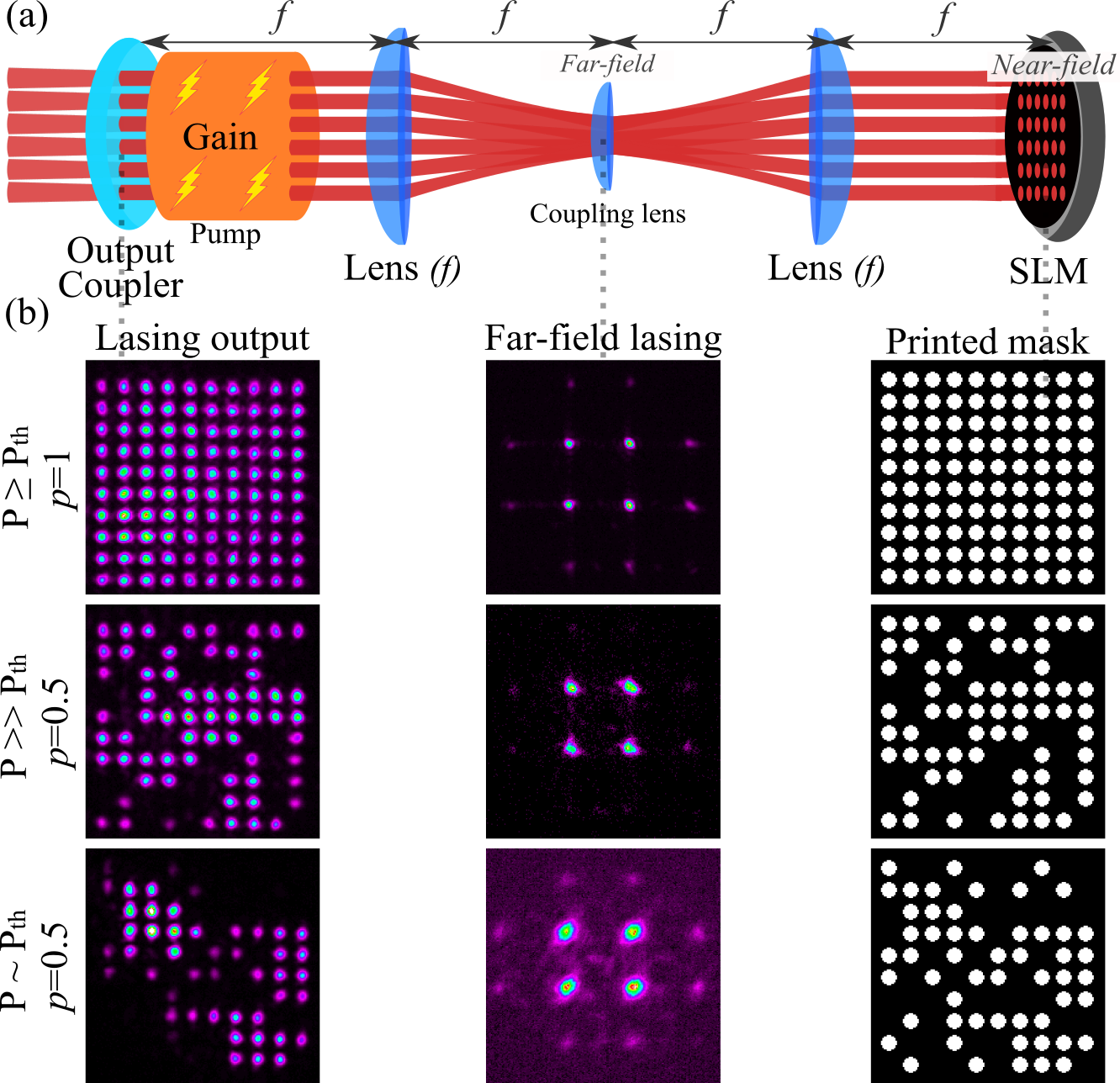}
\caption{Investigating percolation with coupled lasers. (a) Experimental arrangement comprised of a degenerate cavity laser arrangement with a spatial light modulator at one end, serving as laser array generator and a far-field lens, coupling lasers with nearest-neighbors coupling. (b) Lasing near-field output and far-field intensity distributions at high and low pump values.}
\label{fig:10_Exp_sketch_percolation}
\end{figure}

Percolation theory describes a critical transition of a system between a non-connected to a fully connected state. Generally, the system is a network of sites (array), where a site can exchange information with their neighbors depending on the connections state (coupling). Percolation theory is a threshold phenomenon where close to threshold, power-laws with specific exponents are observed for different quantities such as order parameter, correlation length, etc...~\cite{Stauffer94}. Each quantity of the system is characterized by a dedicated exponent. The exponents are related in a universal manner, where only the dimension of the system can change their relation. Each system has a universality class and a scale invariance. Statistical field theory and statistical physics (renormalization group, phase transition, fractals, etc..) have been used for understanding percolation~\cite{Stauffer94}. Due to high number of required degrees of freedom, it has been difficult to perform control experiments for studying percolation. With our recent development and progress in coupling configurations, percolation could be performed with coupled lasers where lasers parameters can be easily controlled. For that, a spatial light modulator at one end of a DCL can serve as a laser array generator and a far-field lens can couple them. In this section, percolation with a square array of coupled lasers is investigated and related to phase-locking of lasers, Fig.~\ref{fig:10_Exp_sketch_percolation}.

Figure~\ref{fig:10_Exp_sketch_percolation}(a) shows the experimental arrangement for investigating percolation with coupled lasers. It is comprised of a DCL arrangement which includes a spatial light modulator (SLM) at one end, a coupling lens at the far-field and an output coupler mirror at the other end. The SLM can spatially modulate a beam of light and can be controlled with a computer, so as to allow a controlled generation of different patterns such as masks for forming array of lasers at different pump pulse realization. A far-field spherical lens, of focal length $f_{C}=f/Z_{a/4}$, mimics the Talbot diffraction at half Talbot distance, and couple lasers with negative nearest neighbors coupling \cite{Mahler19_2}.  

For each realization, each laser (i.e. site) is initialized on/off state following a site probability $p$ (uniformly distributed). Increasing $p$ increases the number of lasers (i.e. the number of sites). The site probability $p$ to have lasers or not in the square array, was controlled by tuning the reflectivity of the SLM. With such control, the intensity distribution at the lasing output, at high pump, is identical to the mask printed on the SLM, as in Fig.~\ref{fig:10_Exp_sketch_percolation}(b). The far-field lasing intensity distribution corresponds to the far-field pattern of out-of-phase locked lasers (i.e. negatively coupled lasers)~\cite{Mahler19_2,Tradonsky17}. The sharpness of the peaks indicates that most of the lasers in the array are phase-locked~\cite{Mahler20_2}.

At low pump, a different behavior is observed where only groups of lasers (clusters) that are relatively large are lasing. A cluster is defined as the group of nearest-neighbors lasers that are all connected to each other. At low pump, the small clusters do not lase. This is due to the fact that mode competition is more severe when the pump value is close to lasing threshold. Due to the strong nearest-neighbors coupling, large clusters, where most of the lasers in the clusters are supported by many neighbors, have a lower loss than small clusters. Thereby, due to amplified mode competition, small cluster are not lasing. 

\begin{figure}[!ht]
\centering
\includegraphics[width=0.6\textwidth]{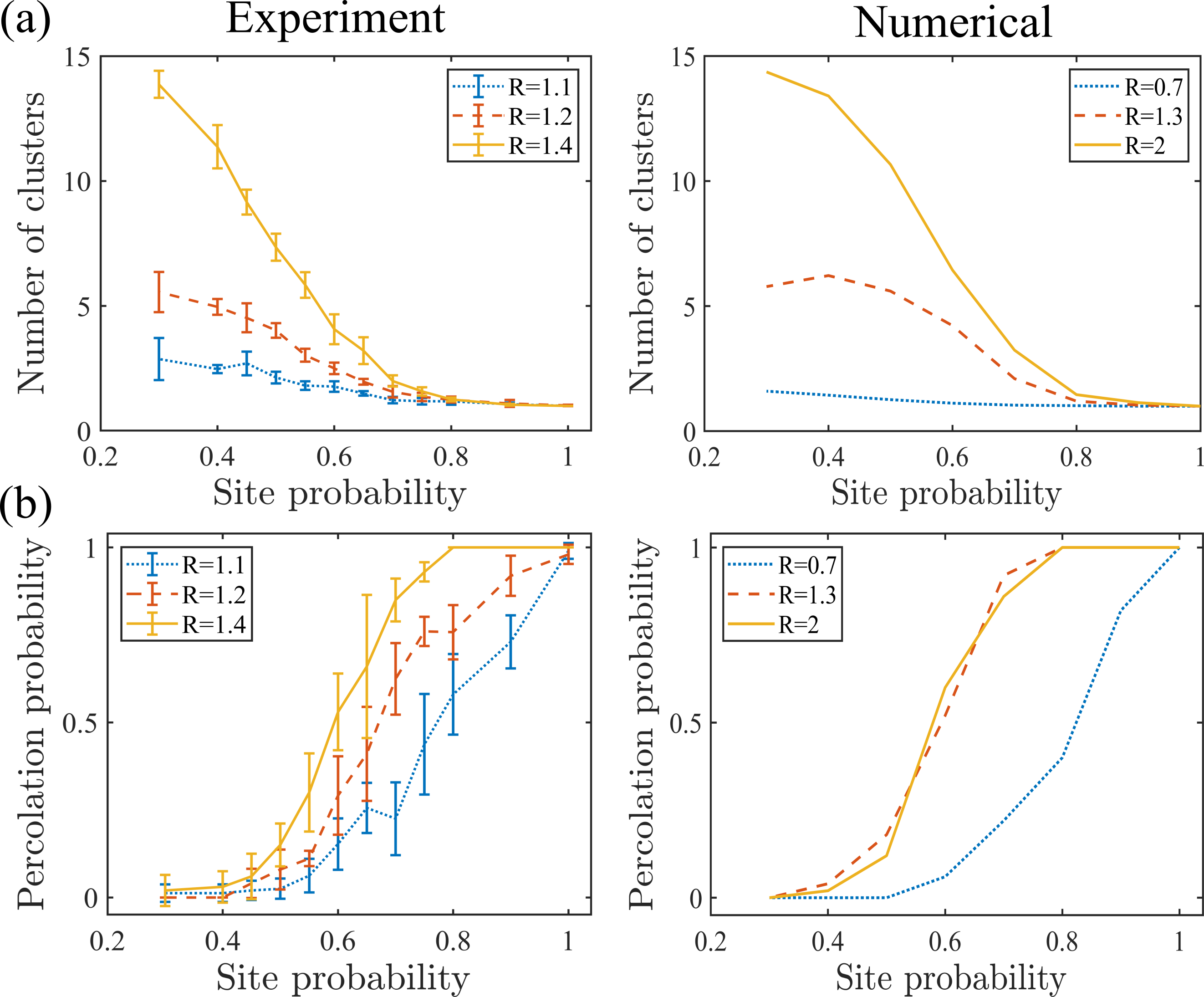}
\caption{Experimental and numerical percolation results with coupled lasers. (a) and (b) Number of clusters and percolation probability as a function of the site probability $p$ for different pump ratio $R$.}
\label{fig:11_PercolationRes}
\end{figure}

Next, the site probability $p$ was varied from $0.3$ to $1$. For each $p$, near-field lasing intensity distributions were recorded for $50$ different mask realizations. The percolation probability was calculated by measuring the percolating state of the array (either $1$ or $0$) at each realization and then taking the ensemble average over the $50$ realizations. The percolating state is defined as $1$ if one can reach both edges of the laser array at $p=1$ by travelling along any cluster in the array, otherwise it is $0$. The number of clusters $N_{c}$ were also measured from the near-field lasing intensity distributions. Results are shown in Fig.~\ref{fig:11_PercolationRes} for different pump ratio $R=P/P_{th}$ values where $R$ is the ratio between the applied pump value $P$ and the pump value at lasing threshold $P_{th}$. 

At high pump ($R=1.4$), the number of clusters, Fig.~\ref{fig:11_PercolationRes}(a), is about $N_{c}=15$ for $p=0.3$ and then decreases as $p$ increases. As shown in Fig.~\ref{fig:10_Exp_sketch_percolation}, all the clusters in the printed mask are lasing at high pump. Thereby the number of clusters corresponds to the number of clusters of the printed mask. 
At lower pump ($R=1.2$), the number of clusters, also decreases but is much lower (about $N_{c}=5$) at $p=0.3$. At even lower pump $R=1.1$, the number is almost $N_{c}=1$ for all $p$, indicating that near lasing threshold, only largest clusters lase, as shown in Fig.~\ref{fig:10_Exp_sketch_percolation}.

The percolation probability, Fig.~\ref{fig:11_PercolationRes}, follows a second-order phase transition as a function of the site probability. However, the transition is sharper and starts at smaller $p$ for higher pump ratios, indicating that near lasing threshold, not only large clusters lase but also some lasers belonging to a large cluster do not lase. This is specifically true for lasers that only have a single nearest neighbors, as they suffer from more loss.  

Our results show that the percolation probability follows a second-order phase transition as a function of the site probability, where the critical probability is around half, as theoretically predicted. We also show that the sharpness of the phase transition and the range of phase-locking both depend on the pump value of the lasers  

\subsection{Fair sampling of XY Hamiltonian \cite{Pal20}}
\label{sec:fairsamp}
In this section, we present a new simulator for the $XY$ spin Hamiltonian based on linearly coupled lasers, that rapidly performs fair sampling by exploiting massive parallelism in the lasers spectral domain \cite{Pal20}. Under the assumption of constant field amplitudes, the coupled lasers are well approximated as Kuramoto phase oscillators \cite{Acebron05}. Then the phases of the lasers can be mapped to the classical XY spins, and the ground state of the classical XY Hamiltonian can be analogous to phase locked steady state of the coupled lasers \cite{Nixon13_3}. Unlike finding the ground state of spin systems by cooling externally (related to the well-known Kibble-Zurek mechanism \cite{Zurek14, Mermin79, Coullet89, Kibble76, Zurek85,Navon15}), in coupled lasers the internal dissipation caused by coupling, drives the lasers into a globally stable phase locked state, identical to the ground state of classical XY spin Hamiltonian \cite{Nixon13_3, Pal17}. In our simulator, each laser has  approximately $250$ longitudinal modes that form an ensemble of approximately $250$ identical but independent simulators of the $XY$ spin Hamiltonian. This provides a massive parallelism that enables rapid and accurate fair sampling of the ground state manifold \cite{Pal20}. We directly measure the statistical average of spin ordering (magnetization) of the ground state manifold by measuring coherence between the lasers in different array geometries having single, double, and many degenerate ground states \cite{Pal20}. 

\begin{figure}[htbp]
\centering 
\includegraphics[width=0.5\textwidth]{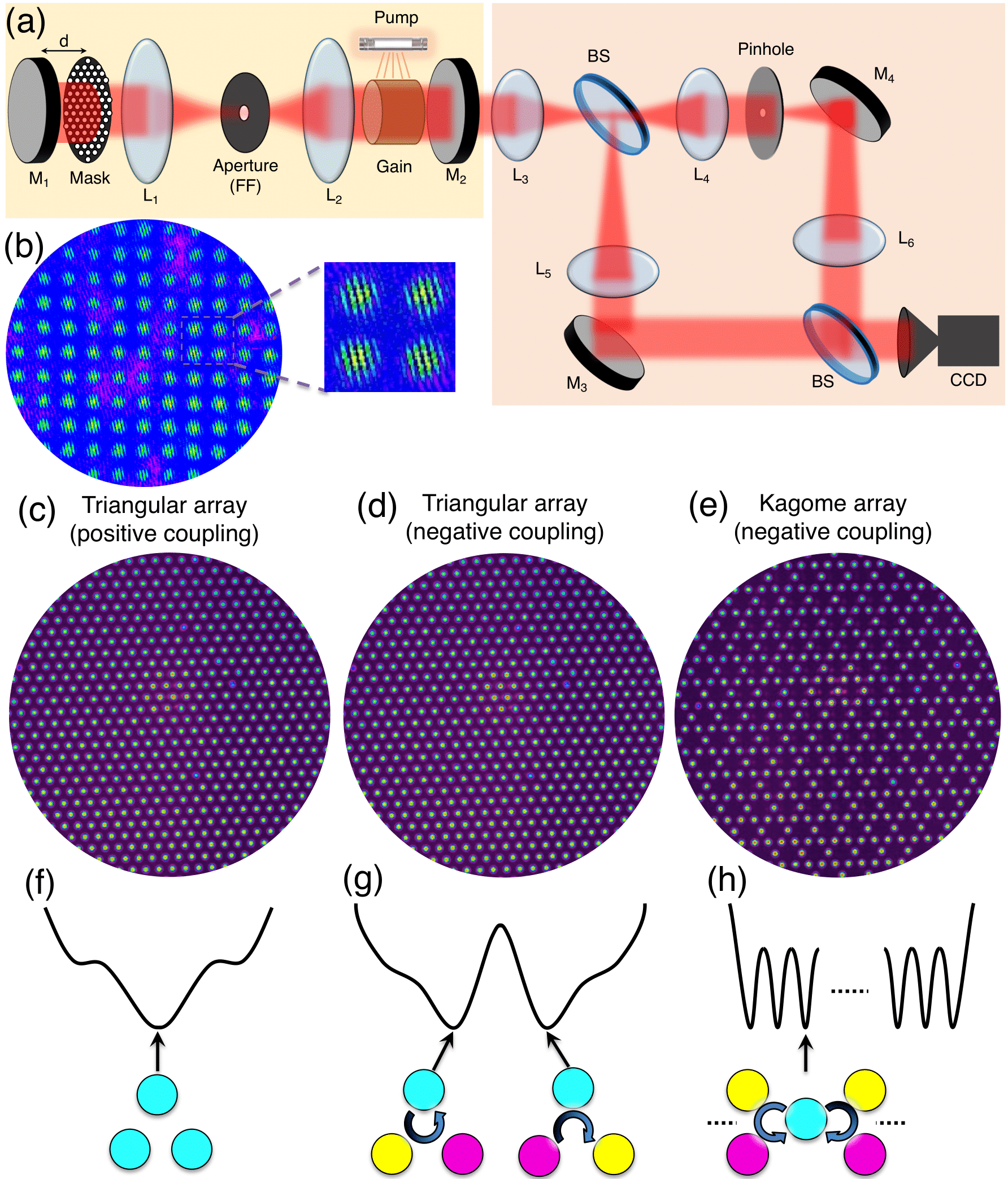}
\caption{Experimental arrangement, array geometries, and representative results for fair sampling of XY Hamiltonian. }
\label{fig:13_XYsamp_exp}
\end{figure}

The experimental arrangement, array configurations and representative results are presented in Fig.~\ref{fig:13_XYsamp_exp}. We performed a series of experiments to demonstrate fair sampling of ground state manifold in square, triangular and Kagome arrays, see more details in Ref.~\cite{Pal20}. 

(a) Schematic of a degenerate cavity laser (shaded in yellow) that forms and phase locks lasers in different array geometries, together with Mach-Zehnder interferometer (shaded in orange) for analyzing the coherence between the lasers. (b) Experimentally measured interference pattern when a single laser interferes with itself and with all the other lasers in the square array. Experimental near-field intensity distributions for (c) triangular array (positive coupling); (d) triangular array (negative coupling); and (e) Kagome array (negative coupling). (f) Landscape with a single ground state, corresponding to in-phase locked triangular array. (g) Landscape with two degenerate ground states, corresponding to vortex and anti-vortex states of out-of-phase locked triangular array. (h) Landscape with highly degenerate ground states, corresponding to $2^{n}$ states ($n$, the number of triangles) in the out-of-phase locked Kagome array. Note, for $n=2$, only $1$ state out of $4$ states is shown. Different colors of the lasers denote different values of the phases. Cyan $=0$; yellow $=2\pi/3$; and purple $=-2\pi/3$. M$_{1}$ and M$_{2}$ denote high reflectivity and partial reflectivity cavity mirrors, M$_{3}$ and M$_{4}$ high reflectivity mirrors. L$_{1}$, L$_{2}$, L$_{3}$, L$_{4}$, L$_{5}$, L$_{6}$ plano-convex lenses, BS beam splitter, and CCD camera.

\subsection{Topological defects dynamics and Kibble-Zurek mechanism \cite{Mahler19}}
\label{sec:KZM}
The evolution of dissipative topological defects and their attraction and annihilation process follows a specific dynamics. Specifically, coupled vortex-antivortex in a square array of coupled oscillators attract and then annihilate each other \cite{Pal17,Mahler19}. They obey a well-defined force that depends on the distance between them. For open boundary conditions, a topological defect can also dissipate (alone) into the boundaries. Then, for small system size, the boundary conditions can influence the dynamics. In this section, using the Kuramoto model, the formation of topological defects is shown to depend on two competing time scales and is related to the Kibble-Zurek mechanism (KZM). The density of topological defects also follows a power law during the dynamics transition as a function of the coupling quench rate, in agreement with the KZM. One critical exponent $\nu$ was extracted, leading to finding the universality class of coupled phase oscillators such as coupled lasers. 

We numerically investigated the temporal evolution of the phase of coupled lasers array by using the Kuramoto model where each laser interacts with its nearest neighbors only and is characterized by a constant amplitude and a time evolving phase \cite{Mahler19}. Since there is no temperature in our Kuramoto model, we define the coupling strength between the lasers $K$ as its analogue (control parameter). Then, uncoupled lasers are equivalent to the infinite temperature state (disordered state) and a complete phase locked state is equivalent to an absolute zero temperature state (ordered state). For an intermediate (not too strong) disorder, when the coupling strength between the lasers is less than a critical value, the lasers are not coupled leading to a fully disordered stable state. When the coupling crosses the critical value, the lasers start to couple, leading to an ordered stable state with some remaining topological defects. 

Finally, when the coupling is much higher than the critical value, the lasers are highly coupled leading to a phase locked (fully ordered) stable state with no topological defects. Ramping the coupling strength in time is then equivalent to varying the control parameter of the system in time, as in KZM. The coupling quench rate determines the slope of the ramp. An infinite coupling quench rate corresponds to a sudden jump in the coupling strength while a zero coupling quench rate corresponds to a constant coupling in time.

\begin{figure}[!ht]
\centering
\includegraphics[width=0.85\textwidth]{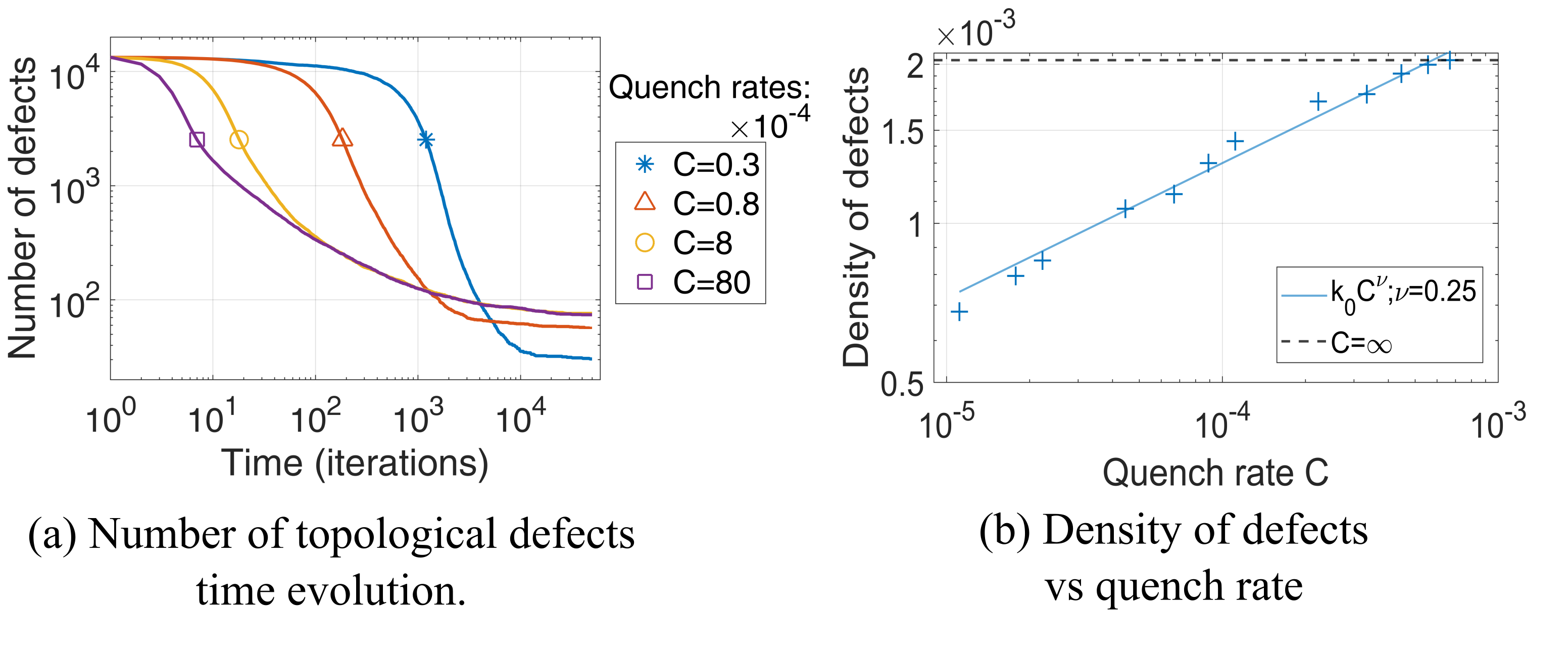}
\caption{The defects dynamics in a periodic square array of $200\times200$ lasers for different coupling quench rates $C$. The results are obtained by averaging over six different realizations, and for each realization the initial conditions of phases and disorder were varied uniformly in the range between $\left[-\pi ~\mathrm{to} ~\pi\right]$ and  $\left[0 ~\mathrm{to} ~\xi=\pi/8\right]$ respectively. (a) The number of defects as a function of time by varying the coupling in time at different rates. (b) Density of defects in the steady-state ($t=35000$) as a function of quench rate $C$. Figure taken from~\cite{Mahler19}.}
\label{fig:16_KZM}
\end{figure}

The formation of dissipative topological defects were observed in coupled phase oscillators and was related to the Kibble-Zurek mechanism (KZM). Their formation depended on two competing time scales and the density of topological defects followed a power law as a function of the coupling quench rate. One exponent $\nu$ extracted from this power law, was $\nu=0.25$, indicating that our coupled phase oscillators systems belong to a new universality class. It is possible to observe the dynamics of topological defects by switching the role of the coupling strength and the disorder. The control parameter (ramped in time) is then the disorder and the coupling strength between the lasers is constant in time, see~\cite{Mahler19} for more details. The same numerical simulations can be done for a one dimensional ring array of lasers, i.e. a chain with periodic boundary conditions where each laser has two nearest neighbors (left and right)~\cite{Mahler19}. Slow variation of the coupling strength might result in lower percentage of topological defects, whereas fast variation results in higher percentage of topological defects, analogue of Kibble-Zurek mechanism in one dimension, see~\cite{Mahler19} for more details. 

\section{Synthetic gauge field \cite{Mahler24_spins}}
\label{sec:Gfield}
\begin{figure}[ht!]
\centering\includegraphics[width=0.65\textwidth]{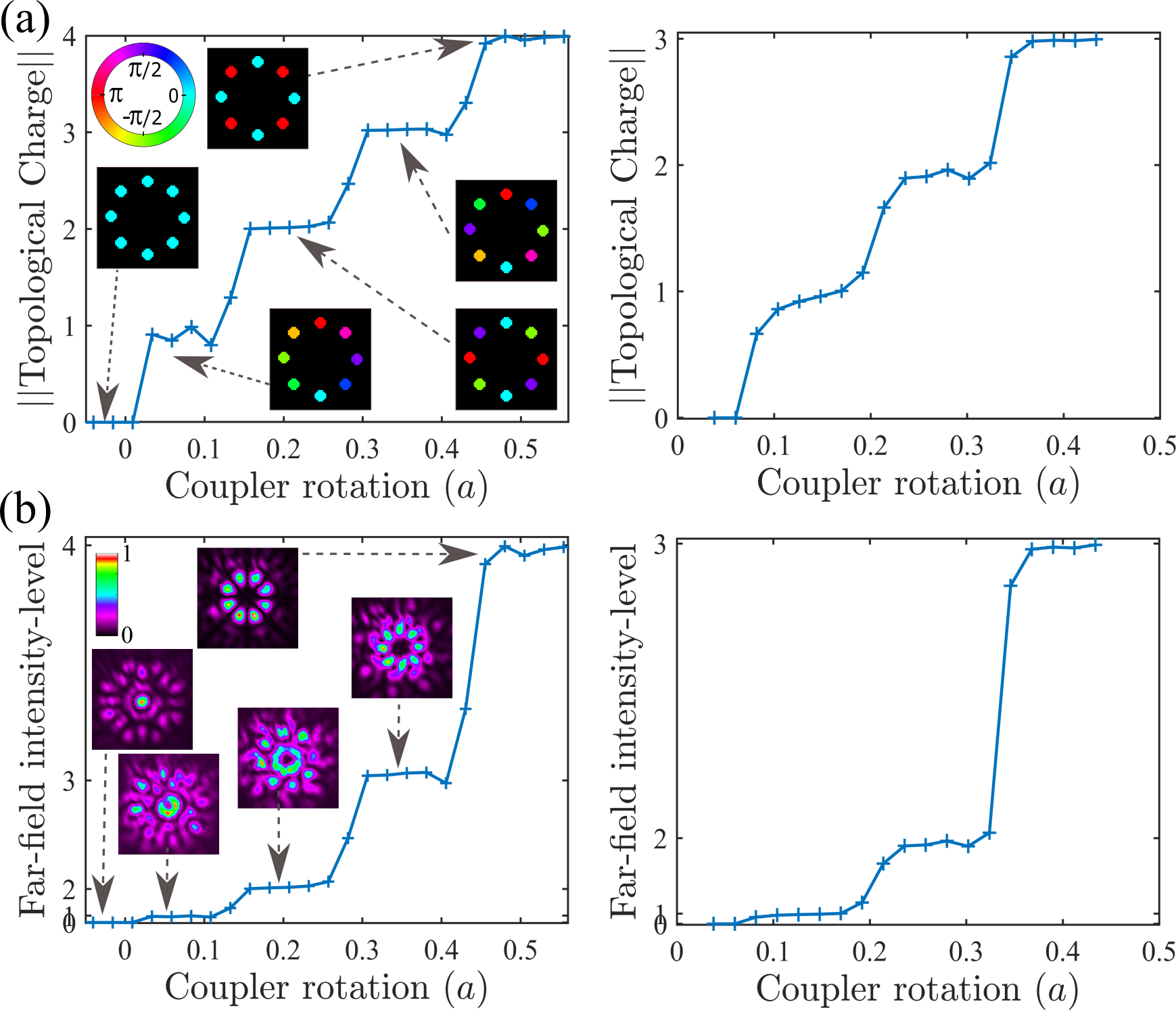}
\caption{Artificial gauge field in a ring array of lasers. (a) Magnitude of the topological charge in a ring of lasers as a function of the coupler rotation, deduced from the far-field intensity distributions showed in (b). (insets) Corresponding phase distribution of the lasers calculated from the numerics. (b) Measured far-field intensity-level in the ring of lasers as a function of the coupler rotation calculated from the far-field intensity distributions showed in the insets. (left column) Ring of $8$ lasers, (right column) ring of $7$ lasers.}
\label{fig:19_ring_TC_op_A}
\end{figure}

We resorted to anti-symmetrical coupling between lasers by means of a ring degenerate cavity laser for implementing arbitrary mid-field coupling, for breaking time-reversal symmetry and for forming artificial gauge field in lasers. First, we demonstrated arbitrary mid-field coupling of lasers in a square array of lasers in Section~\ref{sec:MFK}, where any desired phase locking state can be obtained. In Fig.\ref{fig:19_ring_TC_op_A}, reversal-time symmetry of the lasers was broken by implementing artificial gauge field in a ring array, see Ref. \cite{Mahler24_spins}. The topological phase defect states of a ring of lasers were observed and a sharp first-order-like transition occurred between the different topological states. Results are presented in Fig.~\ref{fig:19_ring_TC_op_A} by showing the phase locked state of the ring array as a function of the coupler rotation. Coupler rotation of zero corresponds to the rotation where in-phase locking of the lasers is achieved and a rotation of $a/2$ corresponds to the rotation where out-of-phase locking is achieved. 

The phase locked state of the lasers can be characterized either from the phase of the lasers or from the far-field intensity distribution. As shown in Fig.~\ref{fig:19_ring_TC_op_A}, for a ring array geometry of $N$ lasers, there exist only $int(N/2)+1$ (where the $int(N/2)$ function returns the integer value of $N/2$) possible phase locked states, each state having a different topological charge~\cite{Pal20}. The topological charge $T.C$ of $N$ lasers can be calculated from the whole array as~\cite{Pal17, Pal20}: 
\begin{equation}
T.C=\frac{1}{2\pi}\sum_{n=1}^{N}\left\{\phi_{n+1}-\phi_{n}\right\},
\label{eq:5_defect_number_1D}
\end{equation} 
where $\{ \}$ wraps phases in the range from $\left[-\pi ~\mathrm{to} ~\pi\right]$ and $\phi$ is the averaged phase of the laser number $n$ in the range from $\left[1 ~\mathrm{to} ~N\right]$. A topological charge of zero $T.C=0$ corresponds to the in-phase locked state, $T.C=\pm4$ corresponds to the out-of-phase locked state and all the other topological charges correspond to intermediate phase locked states. Figure~\ref{fig:19_ring_TC_op_A}(a) shows the magnitude of the topological charge calculated from the far-field intensity distributions (shown in Fig.~\ref{fig:19_ring_TC_op_A}(b)) as a function of the coupler rotation in a ring array of $N=8$ (left) and $N=7$ (right). The insets show the theoretical phase distribution of the lasers corresponding to the topological charge level. 

As evident, by rotating the coupler, the magnitude of the topological charge gradually increases with a sharp first-order-like transition between different charges where th e level of the topological charge increases from $0$ to $int(N/2)+1$. In some transitions, an intermediate point between two plateau levels was measured. This is surprising since all the $int(N/2)+1$ phase-locked states are measured and each state corresponds to a plateau level in Fig.~\ref{fig:19_ring_TC_op_A}. We explain these intermediate points as follow: since our DDCL is composed of many independent longitudinal (temporal) modes~\cite{Mahler21}, then each measurement performed to record the results shown in Fig.~\ref{fig:19_ring_TC_op_A} corresponds to an averaged measurement over hundreds of longitudinal modes. As each longitudinal mode in our DDCL is independent~\cite{Mahler21}, in the presence of loss-degenerate phase-locking state, a superposition of phase-locked states can be measured, as in~\cite{Mahler20}. Thereby, the intermediate point corresponds to a superposition of phase-locked states and can be suppressed by generating a single longitudinal mode DDCL or by coupling the longitudinal modes in the DDCL~\cite{Mahler20}. Except for the intermediate points, the phase-locked states in Fig.~\ref{fig:19_ring_TC_op_A} exhibit stable and minimal-loss phase features that are robust over external detuning or disorder as they always exhibit a single phase-locked state, even in the presence of hundred of independent longitudinal modes.   

Figure~\ref{fig:19_ring_TC_op_A}(b) shows the far-field intensity-level as a function of the coupler rotation. The far-field intensity-level was measured by calculating the dark area present in the center of the far-field intensity distributions (Fig.~\ref{fig:19_ring_TC_op_A}(b) insets). As evident, by rotating the coupler, the intensity-level gradually increases with a sharp first-order-like transition between different levels. These results show that the dissipative coupling ensures stable and robust lasing by selecting the lowest-loss state and removing superposition of states, and that the dispersive coupling ensures the formation of discrete intensity-levels, where the lasers exhibit different topological charge levels for different intensity-levels. Our results indicate that by tuning the pump energy of the lasers, the lasers will exhibit discrete topological charge levels for different pump energies (i.e. pump frequencies), as in the numerical simulations~\cite{Umucal11}. Such behavior is analogous to the Harper-Hofstadter physics~\cite{Hofstadter76,Harper1955} and the integer quantum Hall effect~\cite{Ozawa19} where the spectral properties of non-interacting electrons in a array under the influence of a gauge field were described. 

\section{Phase-locking spatial modes and laser speckles reduction applications}
\label{sec:spatmodes}
In this section, we report our investigations on phase-locking spatial modes of the DCL and its applications on laser speckle reduction and high purity beam generation. Section~\ref{sec:Cvsdt} presents the spatio-temporal mode structure of the DCL and how a phase diffuser can break the frequency degeneracy between spatial modes and reduce speckles both at short and long integration times \cite{Mahler21,Eliezer21}. Section~\ref{sec:imgdiff} shows how the DCL can be used to perform imaging through thin scattering media \cite{Chriki21}. Section~\ref{sec:beamshape} shows how the DCL can be used to generate high-resolution arbitrary shaped laser beams with high brightness and tunable spatial coherence \cite{Tradonsky21}.

\subsection{Spatio-temporal modal structure of the DCL and speckles reduction \cite{Mahler21,Eliezer21}}
\label{sec:Cvsdt}
The degenerate cavity laser (DCL) is a specific laser cavity that can support numerous independent lasing modes, useful for many applications. The different lasing modes in the DCL, up to 320,000 spatial modes, were efficiently used for suppressing speckles \cite{Nixon13,Chriki18}. We showed that speckles can also be reduced efficiently in the microsecond timescale where the temporal (longitudinal) modes play a crucial role \cite{Chriki18}. Speckles level can be quantified by calculating the speckle contrast C (also commonly named K):
\begin{equation}
C=\frac{\sqrt{\langle I^2(\vec{r}_i,\tau)\rangle_{i}-\langle I(\vec{r}_i,\tau)\rangle_{i}^{2}}}{\langle I(\vec{r}_i,\tau) \rangle _{i}}
\end{equation}
with $I$ the time trace intensity of the laser light at spatial location $\vec{r}_i$ and with an exposure time $\tau$. We showed that at long integration times, speckle contrast is reduced as $C=\frac{1}{\sqrt{N}}$ with $N$ the number of spatial modes, $C=\frac{1}{\sqrt{M}}$ at short integration times with $M$ the number of temporal (longitudinal) modes and $C\approx1$ at ultrashort integration times \cite{Chriki18}. 

\begin{figure}[ht!]
\centering\includegraphics[width=0.95\textwidth]{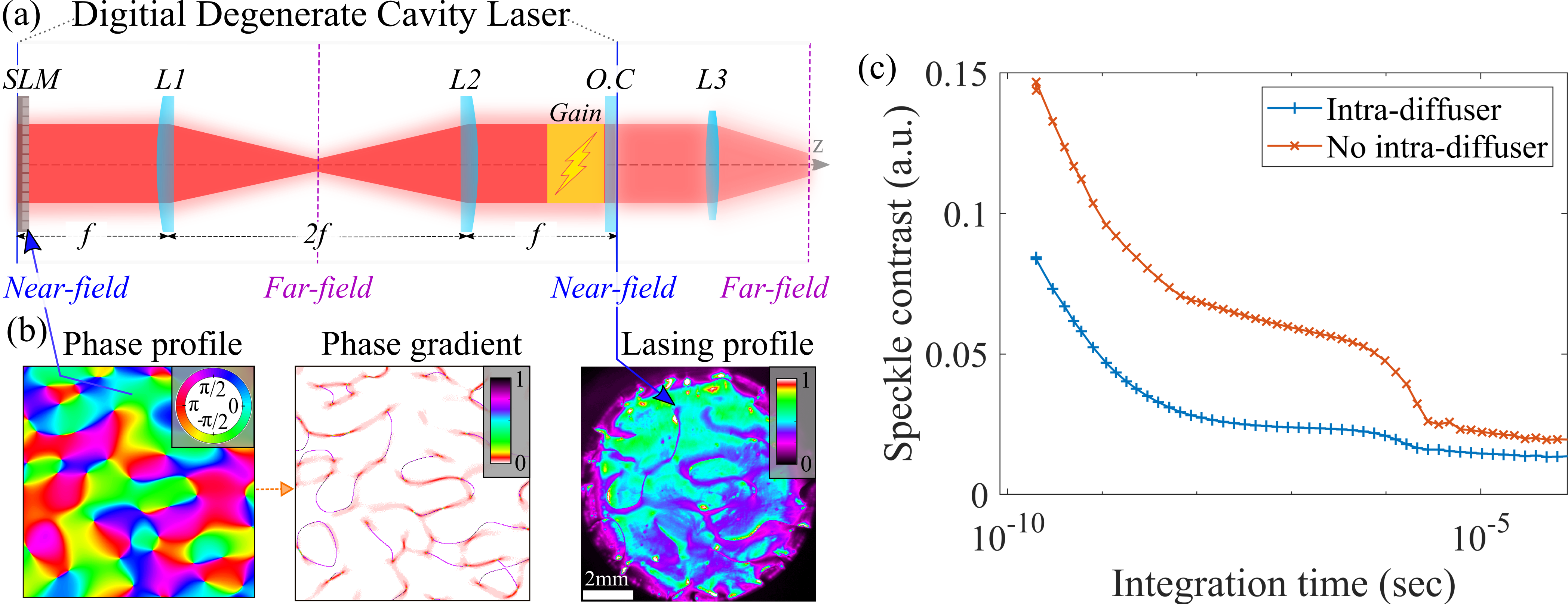}
\caption{Speckles reduction by the means of a DCL and a near-field phase diffuser. (a) Experimental arrangement of a DCL with near-field phase diffuser. (b) Phase profile applied on the SLM, near-field and far-field lasing intensity distributions. (c) Speckle contrast as a function of the integration time. }
\label{fig:20_C_vs_dt}
\end{figure}

In the regime of short timescales, the spatial and temporal modes interact to form spatiotemporal supermodes, such that the spatial degrees of freedom are encoded onto the temporal modes \cite{Chriki18}. In the regime of long timescales, the supermodes are no longer a valid representation of the laser modal structure and the speckle contrast is suppressed according to the number of spatial modes only. Due to this spatiotemporal mechanism, highly multimode lasers can be used for speckle suppression in high-speed full-field imaging applications, as we demonstrated for imaging a fast moving object \cite{Chriki18}.

Recently, we introduced a method to accelerate the spatial decoherence of laser emission, achieving speckle suppression in the nanosecond integration time scale, Fig.~\ref{fig:20_C_vs_dt}. The method relies on the insertion of an intracavity phase diffuser into the near-field of the DCL to break the frequency degeneracy of transverse modes and broaden the lasing spectrum \cite{Mahler21}. Furthermore, we transformed the lasing mode structure of the DCL by means of a random phase profile applied on the near-field plane. By gradually tuning the scaling of the random phase profile, we found a regime where the randomness induces only a relatively low degree of loss, while significantly reducing the gain competition between modes and thereby effectively increasing the number of independent lasing modes. This approach provides a simple and robust method to control the lasing mode structure of the cavity and suppresses the modal competition for gain. The results are presented in this section.

Figure~\ref{fig:20_C_vs_dt}(b) shows an example of a random phase profile applied on the SLM with $\xi$~=~1~mm, the corresponding calculated gradient contours of the phase profile and the detected near-field intensity profile of the DCL with the phase profile. As evident, the transverse intensity profile of the DCL drops abruptly along the high gradient contours of the applied phase profile. Thereby, abrupt variations of the random phase profile reshape the mode structure and segment the beam into multiple localized lasing domains.

To investigate the effects of applying randomness in the transverse dimensions of the cavity, we refer to one-dimensional calculation of the transverse mode structure. In general, a perfectly tuned DCL cannot practically exist, due to alignment imperfections, aberrations and thermal lensing. Therefore, in our mode structure calculation, we incorporate a slight deviation from the purely degenerate case, See Ref. \cite{Eliezer21} for more details. Experimentally, applying a random phase profile in a laser cavity is generally considered deleterious due to diffraction losses. Nevertheless, in agreement with the theoretical prediction for a DCL, we find a regime of sustainable high power emissions even with short phase correlation lengths.

\subsection{Imaging through scattering media \cite{Chriki21}}
\label{sec:imgdiff}
We present an all-optical method for full-field imaging through thin scattering media in real time. It relies on a fundamentally different approach where the modes of a highly multimode laser cavity can phase lock to a minimal loss state by passing only through uniform phase regions of the intra-cavity scattering medium, thereby minimizing scattering losses and ensuring high quality imaging. We show that separating the gain from the scattering media preserves all independent degrees of freedom of the highly multimode laser cavity. This separation enables significant scattering suppression, high quality and high SNR full-field imaging of complex objects through thin scattering media. When the numerical aperture (NA) of the laser cavity is larger than the diffusive angle of the scattering media, the laser selects the modes with minimal round-trip loss, which allows full-field imaging of the object. Due to the exponential buildup of the lasing modes, the time required for the laser to shape the beam traveling inside the cavity can be extremely short, on the order of several round trips~\cite{Chriki18,Mahler21}. Specifically, we show that the buildup time can be as short as $100$ ns in a Q-switched laser, orders of magnitude faster than any other reported wavefront shaping method.

The physical mechanism which enables imaging through scattering media inside a multimode laser cavity is inherently related to the process of mode build-up in lasers. In the pre-lasing stage, photons are spontaneously emitted from the gain medium and due to the random nature of the emission, a large ensemble of possible states (modes) is randomly sampled. These initial modes compete over available gain, until only modes with minimal loss, and their coherent superposition, survive in the steady state lasing stage. Since a self-consistent laser mode must accurately retrace its path after a round-trip in the laser cavity, the minimal-loss modes have the intrinsic property of efficient self-imaging through the intra-cavity scattering media. When many such degenerate low-loss modes exist, their coherent superposition can well approximate an arbitrary image that would then be self-imaged inside the laser cavity. Typical multimode laser cavities support only a small number of lasing modes ($<100$), and consequently are inadequate for optical imaging. Therefore, we resort to the highly multimode DCL arrangement.

\begin{figure}[ht!]
\centering
\includegraphics[width=0.55\linewidth]{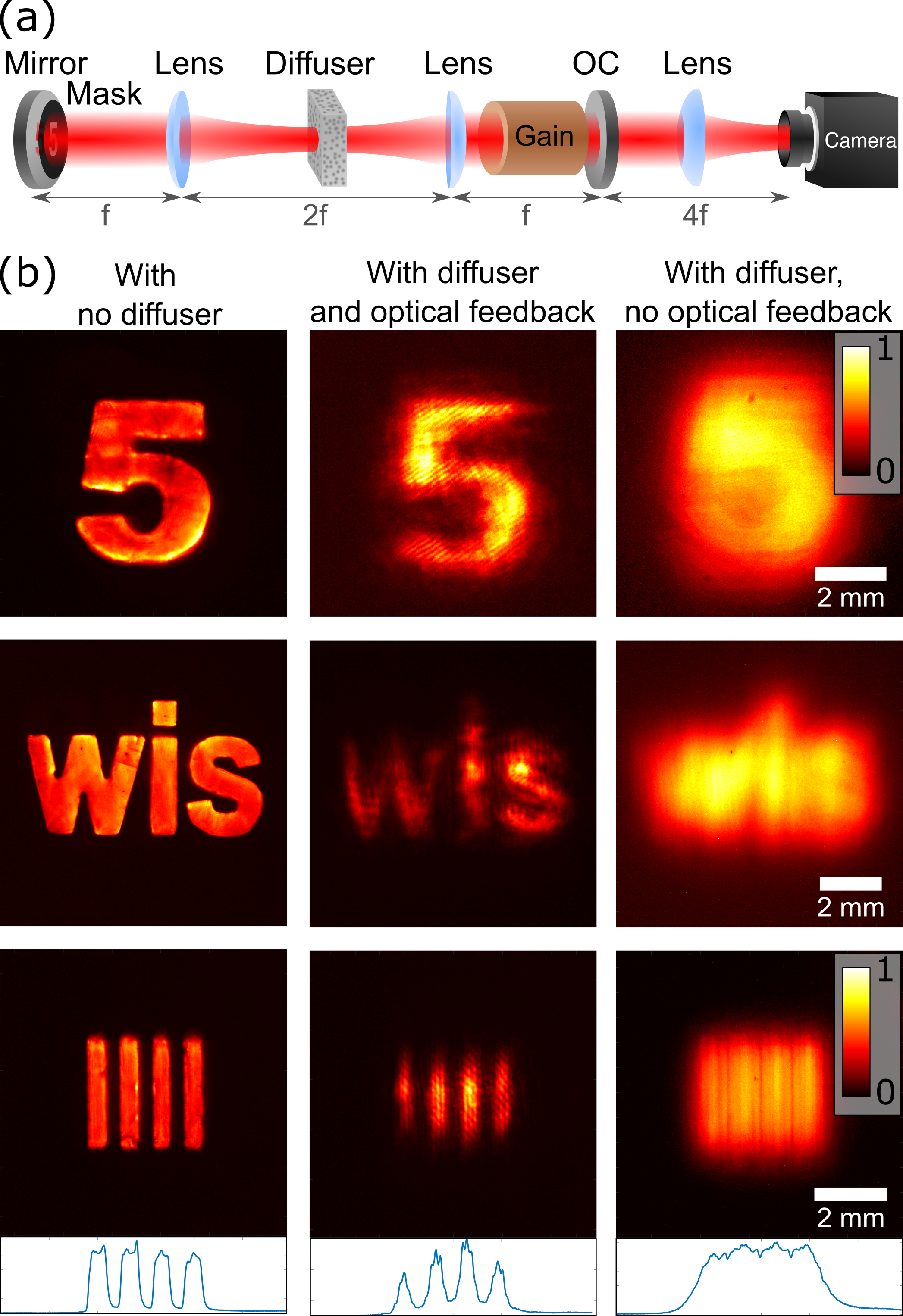}
\caption{Experimental arrangements and results for imaging through weak scattering media. (a) DCL arrangement with a far-field diffuser and a near-field mask to be imaged. The diffuser can also be placed in the mid-field plane (plane between the back mirror and the lens). (b) Intensity distributions detected at the output plane for different mask patterns. (left column) Imaging with a DCL with no diffuser, (middle column) with a DCL with far-field diffuser and (right column) with a $4f$ telescope (incoherent source illumination).}
\label{fig:25_ExpSkcth}
\end{figure}

The experimental DCL arrangement and some typical results are presented in Fig.~\ref{fig:25_ExpSkcth}. In the experimental arrangement, shown in shown in Fig.~\ref{fig:25_ExpSkcth}(a), a binary amplitude mask is placed near the back mirror of the DCL and serves as a reflective object. Figure~\ref{fig:25_ExpSkcth}(b) shows the output intensity distributions without a diffuser, with a diffuser and active lasing configuration, and with a diffuser and passive configuration. When there is no optical diffuser inside the cavity, the mask is imaged onto itself after every round trip, and therefore the lasing output beam is formed in the shape of the mask and clearly identified. When the diffuser is inserted at far-field, the DCL is able to overcome the effect of scattering, and the image of the object is still clearly identified at the output plane of the laser. This is true also in the case where the diffuser is placed at a distance $z_{MF}$ from the back mirror ~\cite{Chriki21}. For comparison, the object masks cannot be identified when using an identical passive $4f$ telescope optical configuration.  

To conclude, a new method for rapid full-field imaging through scattering media, using a highly multimode degenerate cavity laser and intra-cavity diffuser, was presented. Note that the mask and back mirror could be replaced with a single reflective element, and that the method would also work with no optical elements between the object and the diffuser. Such a modified configuration could be suitable for practical applications.

\subsection{High-resolution beam shaping \cite{Tradonsky21}}
\label{sec:beamshape}
In internal beam shaping approaches, separate control of the intensity and coherence were achieved with intra cavity elements. For example, the separate control of the intensity involved the use of intra-cavity spatial light modulator (SLM) in order to digitally select and control desired modes of a conventional (non-degenerate) laser \cite{Ngcobo13}. The separate control of the spatial coherence was done rapidly and efficiently by means of a DCL and an intra-cavity spatial filter aperture and was exploited for wide-field speckle-free illumination and imaging \cite{Nixon13}.

By resorting to a DCL, in which an intra-cavity digital SLM and an intra-cavity spatial Fourier filter are incorporated, it is possible to exploit a very large number (about $10^{5}$) of independent lasing modes with direct access to both the near-field and far-field planes, providing full and independent control of all degrees of freedom of the lasing beam. Based on this approach, we developed a novel, rapid and efficient method to generate high-resolution laser beams with arbitrary intensity, phase and coherence distributions~\cite{Tradonsky21}. To show the efficacy of our method, we generated a variety of unique and high-resolution complex laser beam distributions where the spatial coherence can be rapidly tuned, with little change in the laser output power and by simultaneous control of intensity, phase, and coherence, we demonstrated that a shaped laser beam can be efficiently reshaped after free space propagation into a completely different shape and that extremely high order laser modes can be generated with high fidelity \cite{Tradonsky21}. 

\begin{figure}[ht!]
\centering
\includegraphics[width=0.65\linewidth]{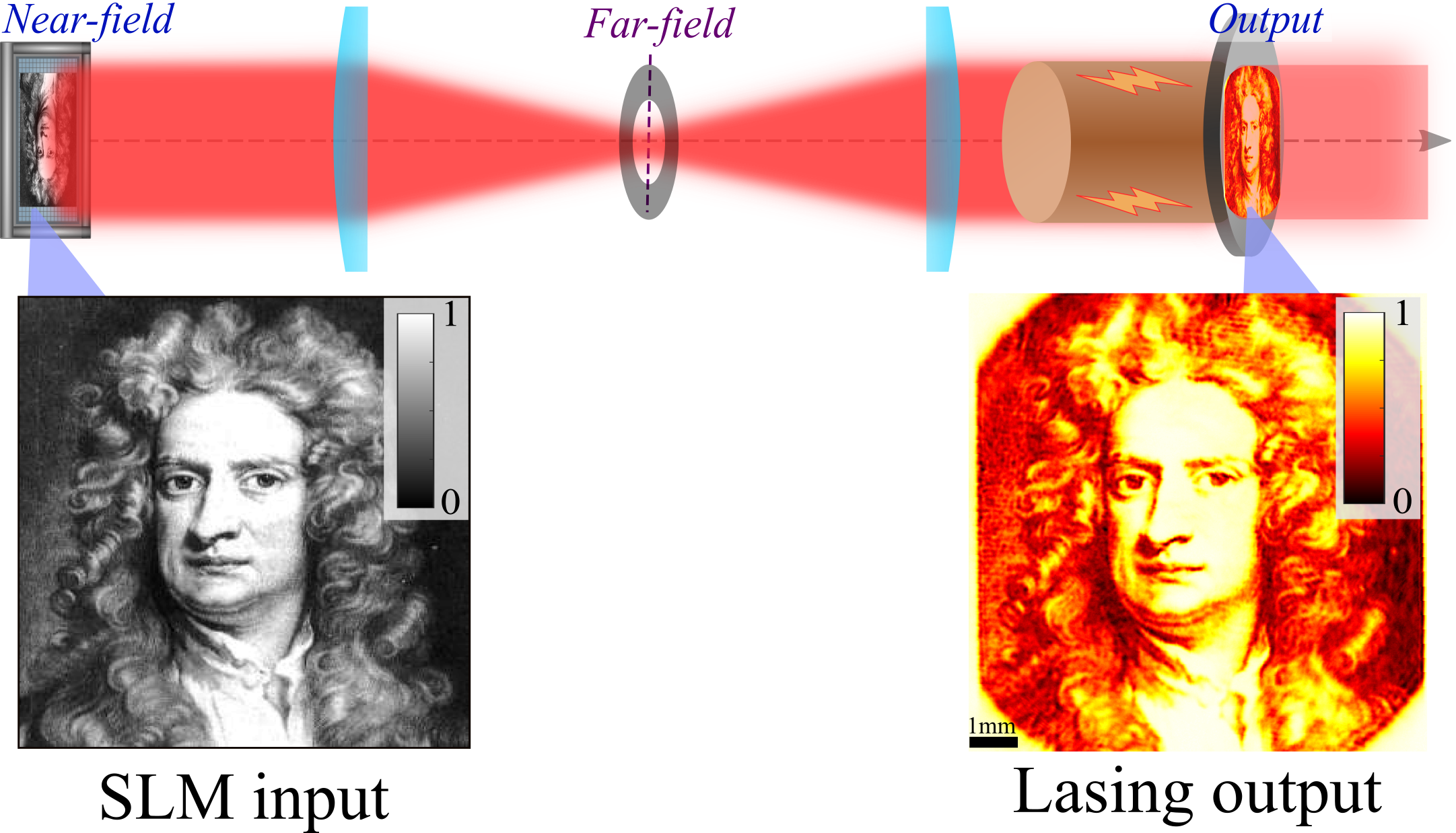}
\caption{Digital DCL arrangement and the generation of a Newton lasing mode (formed in less than $1$ $\mu$s). A laser output beam with a desired distribution is obtained by controlling the intensity, phase, and coherence distributions inside the laser cavity by means of a near-field digital SLM and an adjustable far-field aperture. Newton portrait: © National Portrait Gallery, London.}
\label{fig:30_Skecth_beamshape}
\end{figure}

Figure~\ref{fig:30_Skecth_beamshape} schematically shows the linear digital DCL arrangement. At one end, a reflective phase only SLM serves as a back mirror. At the other end, a ND:YAG gain medium is adjacent to an output coupler. The $4f$ telescope precisely images the field distribution at the output coupler plane onto the SLM plane and vice versa. A laser output beam with a desired distribution is obtained by controlling the intensity, phase, and coherence distributions inside the laser cavity by means of a digital SLM at the near-field and an adjustable intra-cavity aperture at the far-field. 

The control of the reflectivity distribution of the SLM allows control of the intensity distribution of the laser beam in the near-field, whereas control of the phase distribution of the SLM allows control of the frequency distribution. Each super-pixel acts as an independent mirror and all the pixels together form an array of incoherent and independent lasing super-modes~\cite{Eckhouse08}. By varying the size and shape of the far-field aperture, it is possible to control the phase distribution and coherence of the laser light~\cite{Mahler21,Cao19}.

With such arrangement, a high-resolution Newton beam was generated, as shown in Fig.~\ref{fig:30_Skecth_beamshape}. The very high image resolution resulted from the large number of independent lasing modes in the DCL, which can reach up to 320,000~\cite{Nixon13}. With such a large number of independent lasing modes, the resolution is, in principle, limited by the resolution of the SLM. High above lasing threshold, the lasing power of each super-pixel is proportional to its reflectivity, hence the intensity distribution of the laser beam is nearly identical to the image displayed on the SLM (insets). 

\begin{figure}[ht]
\centering
\includegraphics[width=0.55\linewidth]{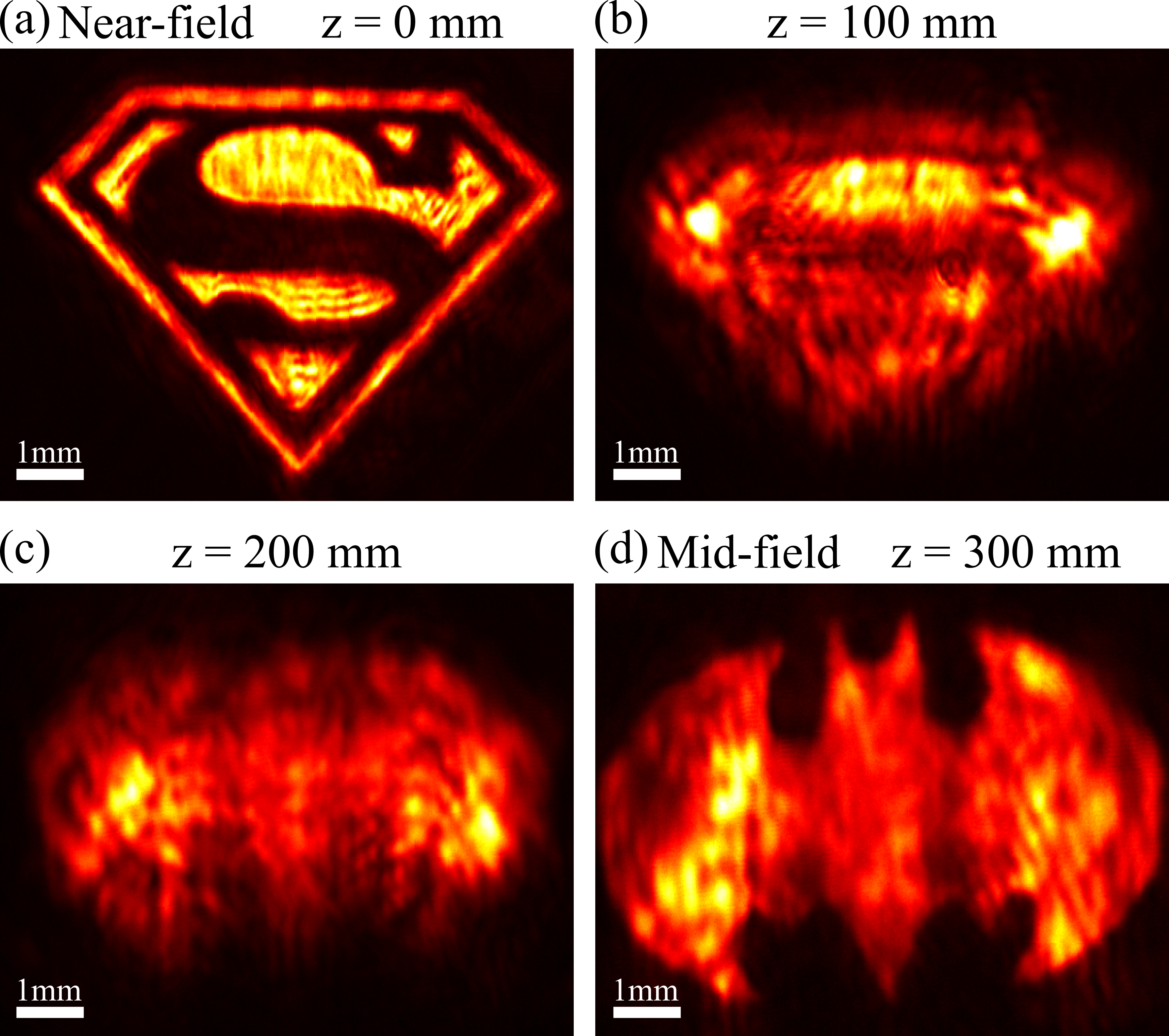}
\caption{Intensity distributions at different propagation distances of a lasing beam with two different intensity distributions at two different propagation distances. (a) Superman image at near-field plane, $z=0$ mm, (b) distorted image at $z=100$ mm, (c) distorted image at $z=200$ mm and (d) Batman image at mid-field plane, $z=300$ mm. }
\label{fig:32_Superman_to_Batman}
\end{figure}

We also obtained a lasing beam with two different intensity distributions at two different propagation distances. For that, we imposed one desired intensity distribution at the near-field plane ($z=0$ mm) by controlling the reflectivity at the intra-cavity SLM, and we inserted an additional shaped aperture at a mid-field plane, located at a distance $z_{MD}$ away from the near-field plane. Typical results at different propagation distances are presented in Fig.~\ref{fig:32_Superman_to_Batman}. Note that the mid-field aperture could be replaced with a transmissive SLM in order to obtain variable control. 

Due to the Fresnel diffraction, the diffracted pattern at a distance $z_{MD}$ away from the near-field plane depends both on the intensity and on the phase distributions at the near-field. While the near-field intensity distribution is constrained by the reflectivity of the SLM, the laser is free to choose the phase distribution that minimizes the loss imposed by the mid-field aperture, i.e. to generate at the mid-field plane, a shape consistent with the mid-field aperture. Thereby, the laser essentially solves a loss minimization problem, whose solution (phase distribution) ensures lasing via both the near-field and mid-field arbitrary intensity distributions. As long as there exist a solution, where loss is smaller than the gain of the laser cavity, lasing is enabled. 

We believe that our method and results sufficiently indicate that it is now possible to generate arbitrary laser output distributions, which could lead to new and interesting applications. Such shaped laser beam can be generated within less than $1$ $\mu$s and with less than $15\%$ reduction in output power~\cite{Tradonsky21}. 

\section{Concluding remarks and future prospects}
\label{sec:conclusion}
To conclude, coupled laser arrays have interesting and rich physical properties that can be exploited to study diverse complex phenomena. During my PhD research, I investigated different coupling methods for improving the phase-locking of arrays of lasers. These methods included nonlinear (spatio-temporal) coupling between lasers by the means of a intra-cavity far-field saturable absorber, that could improve the sensitivity of the lasers to loss differences and for finding unique solution to problems that have several near-degenerate solutions~\cite{Mahler20}, Gaussian coupling between lasers by the means of a far-field Gaussian aperture, so as to obtain steady in-phase locking of laser arrays, regardless of the array geometry, position, orientation, period or size, and even in the presence of near-degenerate solutions, geometric frustration or superimposed independent longitudinal modes~\cite{Naresh21}, far-field coupling between lasers by the means of a spherical lens, cylindrical lens or other optical elements such as diffractive optical element so as to obtain versatile couplings~\cite{Mahler19_2} and mid-field coupling between lasers by the means of a mid-field mask ~\cite{Mahler24_spins}, useful for arbitrary coupling of lasers. 

Using coupled lasers, I also investigated phase transitions and topological defects that occur in several physical phenomena. These included crowd synchrony where a first-order transition to synchronization was observed~\cite{Mahler20_2}, percolation with coupled lasers where a second-order phase transition was observed, fair sampling of XY Hamiltonian, even when frustrated ~\cite{Pal20}, Kibble-Zurek mechanism and dynamics of topological defects in the Kuramoto model~\cite{Mahler19} and synthetic gauge field useful for investigating topological defects~\cite{Mahler24_spins}.

Finally, I investigated phase-locking of spatial modes in the DCL and demonstrated that an intra-cavity phase diffuser can break the frequency degeneracy between spatial modes and further reduces speckles both at short and long integration times \cite{Mahler21,Eliezer21}, demonstrated high quality imaging through thin scattering media \cite{Chriki21} and demonstrated how to generate high-resolution arbitrary shaped laser beams with high brightness and tunable spatial coherence \cite{Tradonsky21}.

Such methods and investigations presented in this thesis can also be implemented in different systems such as optical phased arrays \cite{Heck17}, or side-pumped or end-pumped DCL, or VCSEL arrays \cite{Orenstein92} that are more advanced and integrated systems suitable for industrial applications. As the pump can be easily controlled, new configurations and couplings can be explored including topology, Hofstadter physics, band structures, or phase transitions with disorder coupling. Spatio-temporal control of the many lasing modes in the DCL allow for fast speckle reduction and high-purity laser beam generation with tunable coherence, ideal for laser speckle imaging situations requiring tunable laser sources~\cite{Dong24,HuangJBO_24,Mahler23}. 

\section*{Short Biography}
\label{sec:postdoc}
Simon Mahler received his bachelor degree in physics from Aix Marseille University and University Paris-Sud in 2015. He received his MSc degree in physics of complex systems from University Paris-Saclay in 2017 along with a Magistère in fundemental physics from Unversity Paris-Sud Orsay. During his Magistere's intership, he tested Dubin's model for efficient number diagnostic of long ion chains \cite{Kamsap2017}. During his master's thesis, he investigated coupled arrays of lasers \cite{Mahler19} and spatiotemporal laser speckle reduction \cite{Chriki18}. 

Simon Mahler received in 2021 his PhD in nonlinear optics and engineering with Professors Nir Davidson and Asher A. Friesem at the Weizmann Institute of Science. During his PhD, Simon investigated phase transitions with coupled lasers \cite{Roadmap21,Pal20,Chriki18,Mahler20,Tradonsky21,Mahler21,Eliezer21,Mahler20_2,Naresh21,Mahler19_2,Davidson22,Mahler19,Mahler24_spins,Reddy_2019}. 

Since January 2022, he is a postdoctoral research associate in the Biophotonics Lab of Professor Changhuei Yang at the California Institute of Technology where he is designing non-invasive laser speckle imaging devices for measuring cerebral blood flow on humans 
\cite{Mahler23,Huang23,Zhou24,HuangJBO_24,Dong24,Huang24_stroke,MahlerTBI25,Readhead24}, and imaging vascularized blood vessels in avian embryos \cite{Dong24,Readhead24} for biophotonics applications. 

He is the 2024-recipient of the prestigious SPIE-Franz Hillenkamp Postdoctoral Fellowship. His current research interests include biomedical imaging and engineering, cardiovascular and cerebrovascular monitoring devices, laser speckle imaging, nonlinear optics, spatiotemporal lasing dynamics, and translational research.

{\footnotesize

\printbibliography}

\end{document}